\documentclass[twocolumn]{aastex631}

\usepackage{graphicx,epstopdf,amsmath}
\usepackage{xltabular}
\usepackage{hyperref}

\shorttitle{Switchbacks from Torsional Alfv\'{e}n waves}
\shortauthors{Wyper et al.}
\graphicspath{{./}{figures/}}

\usepackage{color}
\usepackage{amsfonts,amsmath,bm}

\newcommand{\Bb}{\mathbf{B}}

\newcommand{\vb}{\mathbf{v}}

\begin{document}

\title{Magnetic switchback development from the evolution of torsional Alfv\'{e}n waves}

\author[0000-0002-6442-7818]{Peter F. Wyper}
\affiliation{Department of Mathematical Sciences, Durham University, Durham, DH1 3LE, UK}
\email{peter.f.wyper@durham.ac.uk}

\author[0000-0002-2900-0608]{Etienne Pariat}
\affiliation{French-Spanish Laboratory for Astrophysics in Canarias (FSLAC), CNRS IRL2009, Instituto de Astrofísica de Canarias (IAC), 38205 La Laguna, Tenerife, Spain}
\affiliation{Sorbonne Universit\'e, \'Ecole polytechnique, Institut Polytechnique de Paris, Universit\'e Paris Saclay, Observatoire de Paris-PSL, CNRS, Laboratoire de Physique des Plasmas (LPP), 75005 Paris, France}

\author[0000-0002-6276-7771]{Lorenzo Matteini} 
\affiliation{Physics Department, Imperial College London, South Kensington, SW7 2BW, London, UK}

\author[0000-0001-8479-962X]{Jonathan Squire}
\affiliation{Department of Physics, University of Otago, PO Box 56, Dunedin, 9016, NZ}

\author[0009-0003-5268-5128]{Jade Touresse} 
\affiliation{Sorbonne Universit\'e, \'Ecole polytechnique, Institut Polytechnique de Paris, Universit\'e Paris Saclay, Observatoire de Paris-PSL, CNRS, Laboratoire de Physique des Plasmas (LPP), 75005 Paris, France}



\begin{abstract}
Magnetic switchbacks are mostly Alfv\'{e}nic local deflections, or even reversals, of the ambient magnetic field of the solar wind. Despite intense research their origin remains unclear and is highly debated. Here we describe a new simulation study demonstrating how torsional Alfv\'{e}n waves can evolve into switchbacks. We use surface rotational motions to launch torsional Alfv\'{e}n waves into the super-radially expanding magnetic field of a coronal hole. The simulations demonstrate that in such an expanding wind, the waves initially steepen as they propagate while also inducing a cylindrical parallel plasma jet. A combination of shear from the jet and wave evolution under the constraint of approximate spherical polarisation ($|\mathbf{B}|\approx const.$) then leads to the formation of switchbacks in a ring configuration. We explore the 3D nature of the switchbacks, how and where they from, and present scalings for their peak deflection and parallel and perpendicular length scales. We also derive an expression for switchback deflection in the linear regime which we contrast against our results. The results are discussed in relation to recent observations and previous switchback formation studies. 
\end{abstract}

\keywords{Sun: corona; Sun: magnetic fields}


\section{Introduction} \label{sec:intro}
Although their existence was implied by previous missions \citep[e.g.][]{Horbury2018}, an unexpected discovery of Parker Solar Probe was the high prevalence and amplitude of small-scale features known as magnetic switchbacks close to the Sun \citep{Bale2019,Kasper2019}. The inner heliospheric solar wind appears to be dominated by them, yet their origins remain unclear \citep[e.g.][]{wyper2026}. However, their prevalence and energy content suggests they could play an important role in heating and accelerating the solar wind close to the Sun \citep[e.g.][]{Akhavan-Tafti2022,Raouafi2023b}. 

Switchbacks when measured in-situ have a correlated magnetic and velocity perturbation and typically show only small variations in magnetic field magnitude and density \citep[e.g.][]{Bale2019,Horbury2020,Badman2025}. The typical switchback is therefore Alfv\'{e}nic in nature with the perturbation to the magnetic field in the form of a deflection, sometimes to the point of fully reversing the ambient magnetic field direction. Where and if a cut off lies between background Alfv\'{e}nic fluctuations and switchbacks is still unclear \citep[e.g.][]{DudokdeWit2020,Badman2025}. However, any complete switchback formation scenario should be able to produce deflections with a change in sign of $B_R$, i.e. beyond $90^\circ$. In this work we will use the term ``switchback" to refer to perturbations where the deflection angle exceeds $90^\circ$, but note that all of the perturbations in the simulations discussed in this work could be called switchbacks using a more broad definition. 

There are a number of theories for how switchbacks could be formed, see e.g. \citet{wyper2026} for a review. Some theories suppose a coronal origin while others center on mechanisms occurring within the solar wind (in-situ). One prominent in-situ mechanism for switchback formation is the evolution of Alfv\'{e}n waves under the constraint of spherical polarisation within the expanding solar wind, i.e. Alfv\'{e}n waves for which $|\mathbf{B}|$ remains constant as they evolve \citep[e.g.][]{Squire2020,Shoda2021}. Another related mechanism assumes pre-existing cross-field flow or speed gradients acting on Alfv\'{e}n waves \citep{Toth2023}. Both pre-suppose low amplitude Alfv\'{e}nic perturbations entering the solar wind from below which are then amplified, becoming switchbacks as they propagate. To date simulations exploring these mechanisms have been seeded with an isotropic turbulent spectra of fluctuations or simplified wave trains, but so far they have struggled to reproduce the prevalence of observed switchbacks and naturally lack any large-scale structuring or clustering of the switchbacks as is often observed \citep[e.g.][]{Horbury2020,Shi2022}.  

Interchange reconnection, either in a quiescent background form \citep[e.g.][]{Zank2020,Fisk2020,Liang2021} or in an impulsive form as in coronal jets or jet-lets \citep[e.g.][]{Sterling2020,Raouafi2023}, has emerged as a prominent potential coronal origin mechanism. {However, directly linking switchbacks to bright points or jet events has proved challenging due to the errors inherent in mapping from PSP to the solar surface \citep{Bizien2025} with most studies being statistical in nature 
\citep[e.g.,][see also \citeauthor{Tripathy2026} \citeyear{Tripathy2026} for a review]{Kumar2023,Dai2026}. Furthermore,} using 3D MHD simulations we have argued that a key issue with this suggestion is that the reversals in the radial magnetic field component initially present in newly reconnected kinked field lines and plasmoid flux ropes do not survive the ejection process (\citet{Wyper2022,Touresse2024}, see also e.g. \citet{Yang2025}). We note that recently \citet{Longcope2025} presented a switchback formed by solving the MHD equations in the thin flux tube approximation in an idealised configuration representing a newly reconnected flux tube in an interchange context. However, as the authors note the thin flux tube approximation does not account for interaction with neighbouring flux tubes, which is a key reason why switchbacks fail to form in our (and others) 2D and 3D MHD simulations. As we shall discuss though, the wave that is launched has strong similarities to the switchbacks formed in-situ in this work.

Most work therefore suggests that interchange reconnection alone cannot explain switchbacks with deflections above $90^\circ$. Instead in \citet{Wyper2022} we showed that bursty interchange reconnection involving plasmoids launches torsional Alfv\'{e}n waves as the plasmoids are ejected. Furthermore, many coronal jets, and smaller jet-like events involve the transfer of closed-field magnetic twist to the open magnetic field of the coronal hole, again launching torsional Alfv\'{e}n waves \citep{Pariat2009,Lynch2014,Moore2015,Karpen2017,Wyper2017,Wyper2018,Roberts2018,Kumar2018,Kumar2019,Touresse2024,Touresse2026}. As interchange reconnection is integral to this twist transfer, a combination of small and large-scale torsional waves should in fact be produced \citep{Wyper2016b}. Therefore, we argue that interchange reconnection is a major source of impulsive torsional Alfv\'{e}n wave pulses propagating into the solar wind from the corona. 

Aside from interchange reconnection, observations of the low chromosphere also reveal many impulsive swirling motions \citep[e.g.][]{Wedemeyer2012,Giagkiozis2018}. Such swirls are also thought to launch torsional Alfv\'{e}n waves into the high chromosphere with a fraction of them escaping into the corona \citep[e.g.][]{Iijima2017,Liu2019b,Finley2022,Kuniyoshi2025,Lee2025}. However, although such waves can result in magnetic deflections beyond $90^\circ$ in the chromosphere \citep[e.g.][]{Magyar2021}, these deflections are destroyed by the rapid increase in Alfv\'{e}n speed across the transition region. The result is that torsional Alfv\'{e}n wave pulses generated by swirls end up with long wavelengths, and low magnetic deflections as they enter the corona. 

There are therefore a multitude of torsional Alfv\'{e}n wave pulses of differing durations, wavelengths and energies entering the solar wind from below \citep{Morton2026}. A natural question to ask is whether these torsional Alfv\'{e}n waves could act as seeds for in-situ mechanisms to evolve them into switchbacks? And if so, what is the structure of the resulting switchbacks and how does it depend on the properties of the initial wave?

Motivated by these questions we have modeled the evolution of single pulses of torsional Alfv\'{e}n waves as they propagate from the low corona into the solar wind. A key new feature of our simulations is that the solar wind field is taken from a full-sun MHD solar wind quasi-equilibrium. The flux tube along which the waves propagate has an expansion factor of $\sim 11$, exceeding previous global simulations using a radial field (with an expansion factor of $1$ by definition, e.g. \citet{Shoda2021,Touresse2024}). We find that switchbacks form readily as the Alfv\'{e}n waves evolve in our simulation and we describe their properties. In \S\ref{sec:review} we briefly review some key results from the study of Alfv\'{e}n waves in the solar wind that will be relevant to our discussion. \S\ref{sec:setup} describes the simulation setup, \S\ref{sec:example} presents the detailed evolution of a single wave, while \S\ref{sec:para} explores the parameter dependence of the switchbacks on the initial properties of the waves. Finally, in \S\ref{sec:discussion} and \S\ref{sec:summary} we discuss our findings and present our conclusions.

\section{Evolution of Alfv\'{e}nic perturbations in an expanding wind}
\label{sec:review}
Before presenting the simulation results, it will be instructive to summarise some key features of how Alfv\'{e}nic perturbations are thought to evolve into switchbacks in an expanding solar wind as suggested by e.g. \citet{Squire2020,Matteini2024}. This is based upon two key observations. The first is that Alfv\'{e}nic perturbations of \emph{any amplitude} are solutions to the non-linear ideal MHD equations. Specifically, when the flow and magnetic field vectors are decomposed into a background and perturbation field such that
\begin{equation}
\mathbf{B} = \delta \mathbf{b} + \mathbf{B}_0, \quad \mathbf{v} = \delta \mathbf{v} + \mathbf{v}_{sw},
\end{equation}
where we assume here that $\mathbf{B}_0$ points outwards. When the density is assumed constant within the structure so that $\rho = \rho_0$, then perturbations with
\begin{equation}
\delta \mathbf{v} = -\frac{\delta \mathbf{b}}{\sqrt{\mu_0 \rho_0}},
\label{eqn:delta}
\end{equation}
represent outward propagating equilibria of the momentum and induction equation in a moving reference frame of speed 
\begin{equation}
v_{ref} = v_a + v_{sw},
\end{equation}
where $v_a = \frac{|\mathbf{B}_0|}{\sqrt{\mu_0 \rho_0}}$ is the Alfv\'{e}n speed of the background magnetic field -- the so-called deHoffmann-Teller frame \citep{deHoffmann1950}. This is provided that \citep[e.g.][]{Barnes1974,Hollweg1974},
\begin{equation}
P + \frac{|\mathbf{B}|^2}{2\mu_0} = const.
\end{equation}
In the moving frame of reference the total velocity in the outward moving wave is parallel to the total magnetic field, with the fields of each related by 
\begin{equation}
\label{eqn:vb}
\vb =  -\frac{\Bb}{\sqrt{\mu_0 \rho_0}}. 
\end{equation}
Although pressure balanced equilbria are possible, generally switchbacks tend to have constant pressure \citep[$P$, e.g.][]{Badman2025} and therefore typically the magnetic field within switchbacks is constrained by
\begin{equation}
|\mathbf{B}| = |\delta \mathbf{b} + \mathbf{B}_0| = const.,
\end{equation}
so that the magnetic field vector can take any direction so long as it stays on the surface of a sphere defined by $|\mathbf{B}|=const.$, i.e. the wave is spherically polarised. Such equilibria on their own are relatively stable in uniform background fields but eventually reduce in amplitude via for example the parametric decay instability \citep[e.g.][]{Tenerani2020,Shi2024}. However, the second key observation is that Alfv\'{e}n waves propagating in expanding magnetic flux tubes grow in amplitude as the wave speed ($v_{sw}+v_a$) drops with distance. As a result, $|\delta \mathbf{b}|/|\mathbf{B}_0|$ increases as the wave propagates. Combined with the desire to maintain spherical polarisation so as to stay locally in force balance, small amplitude waves can effectively evolve through a series of equilibria into large amplitude perturbations. \citet{Mallet2021} showed that one way this can be achieved is by a coupling to second order compressive fluctuations. In the extreme $|\delta \mathbf{b}|/|\mathbf{B}_0|$ can eventually become greater than one and the field develops a reversal in the radial field component, i.e. becomes a large-amplitude switchback.

\begin{figure*}
\centering
\includegraphics[width=\textwidth]{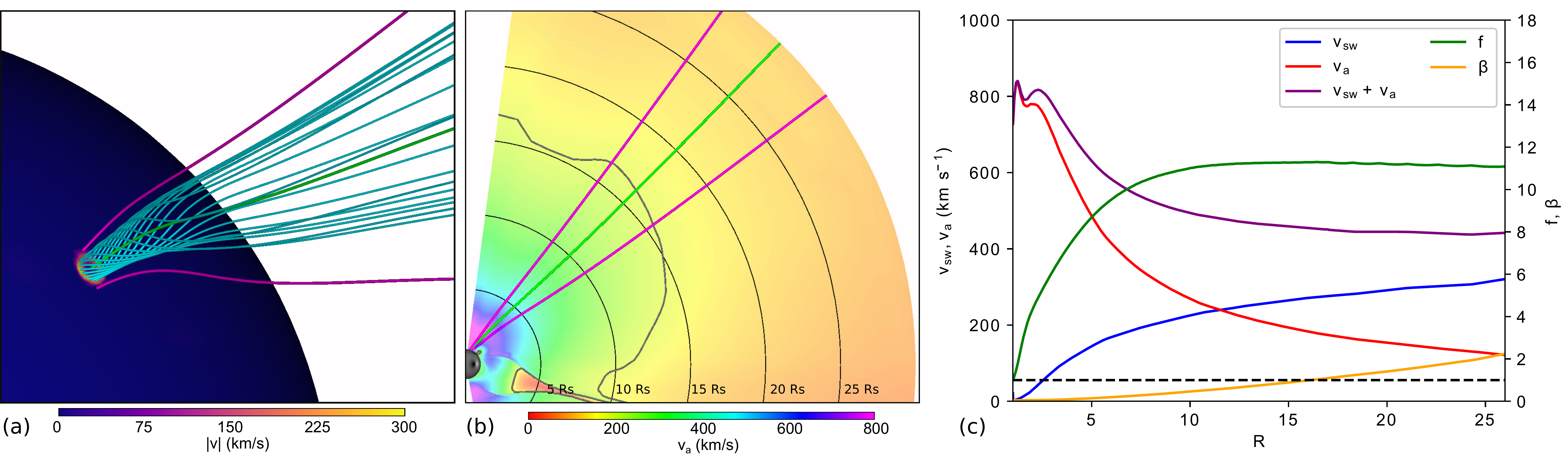}
\caption{(a) Surface shading shows the rotational surface driving (for case 10). Cyan field lines show the twist within the torsional Alfv\'{e}n wave launched by the driving (at $t=750$\,s). The green field line shows the central axis of the rotation. Pink field lines are traced from the edge of the driving region, highlighting the edge of the flux tube within which the wave propagates. (b) $v_a$ in a plane of constant $\phi$, showing the Alfv\'{e}n speed variation along the flux tube. The green and pink field lines are the same as in (a). The dark grey contour shows where $\beta=1$. (c) Solar wind speed ($v_{sw}$, blue line), Alfvén speed ($v_a$, red line), expansion factor (f, green line), and plasma beta ($\beta$, orange line), Alfvén wave propagation speed ($v_{sw}+v_a$, purple line), measured along the green axis field line. } 
\label{fig:setup}
\end{figure*}

\begin{table}
\begin{center} 
\caption{Overview of the simulations, grouped according to peak number of turns injected by the surface driving profile.}
\label{tab}
\begin{tabular}{ |c|c|c|c| } 
 \hline
 Case & $(v_{drive})_{max}$ (km\,s$^{-1}$) & $t_{period}$ (s) & no. turns \\ 
 \hline
 1 & 604 ($\approx 0.8v_a$) & 300 & 0.6 \\ 
 2 & 604 ($\approx 0.8v_a$) & 450 & 0.9 \\
 3 & 604 ($\approx 0.8v_a$) & 600 & 1.2 \\
 4 & 604 ($\approx 0.8v_a$) & 750 & 1.5 \\
 5 & 604 ($\approx 0.8v_a$) & 900 & 1.8 \\
 6 & 604 ($\approx 0.8v_a$) & 1050 & 2.1 \\
 7 & 604 ($\approx 0.8v_a$) & 1200 & 2.4 \\
\hline
 8 & 302 ($\approx 0.4v_a$) & 600 & 0.6 \\ 
 9 & 302 ($\approx 0.4v_a$) & 900 & 0.9 \\
 10 & 302 ($\approx 0.4v_a$) & 1200 & 1.2 \\
 11 & 302 ($\approx 0.4v_a$) & 1500 & 1.5 \\
 12 & 302 ($\approx 0.4v_a$) & 1800 & 1.8 \\
 13 & 302 ($\approx 0.4v_a$) & 2100 & 2.1 \\
 14 & 302 ($\approx 0.4v_a$) & 2400 & 2.4 \\
 \hline
 15 & 151 ($\approx 0.2v_a$) & 1200 & 0.6 \\ 
 16 & 151 ($\approx 0.2v_a$) & 1800 & 0.9 \\
 17 & 151 ($\approx 0.2v_a$) & 2400 & 1.2 \\
 18 & 151 ($\approx 0.2v_a$) & 3000 & 1.5 \\
 19 & 151 ($\approx 0.2v_a$) & 3600 & 1.8 \\
 20 & 151 ($\approx 0.2v_a$) & 4200 & 2.1 \\
 21 & 151 ($\approx 0.2v_a$) & 4800 & 2.4 \\
 \hline
\end{tabular}
\end{center}
\end{table}

\section{Simulation Setup}
\label{sec:setup}
The simulations are initialised with the relaxed quasi-equilibrium used in \citet{Wyper2021}. Here we only summarise the key features of the initial condition and point the reader towards this paper for more specifics. The spherical simulation domain extends out to $30$ solar radii ($R_\odot$), where open free-slip boundary conditions are imposed. Note that all radial distances in this work are heliocentric. The domain includes $\phi \in \pm 180^\circ$, and extends between $\pm 81^\circ$ in $\theta$ to avoid singularities at the poles. Open, free-slip boundary conditions are also imposed on these side boundaries. The magnetic field in the volume starts as an idealised potential field source surface magnetic field containing an equatorial active region, helmet streamer belt, pseudostreamer and a low latitude coronal hole. Superimposed upon this magnetic field is an isothermal 1D Parker solar wind solution \citep{Parker1958}. This was then allowed to evolve into the quasi-equilibirum initial condition we use here in which the coronal hole wind has become field aligned and the heliospheric current sheet contains plasmoids that are quasi-periodically ejected. The end of this relaxation phase is taken as the initial reference time, $t=0$, of the simulations presented in this study. The grid is adaptively refined, with four levels of additional refinement above the background. The compressible ideal MHD equations are solved on this grid using the Adaptively Refined Magnetohydrodynamic Solver \citep[ARMS;][]{DeVore2008}, under the assumption of an ideal gas and constant temperature ($T=1\,MK$), consistent with the imposed Parker wind \citep[][]{Karpen2017,Wyper2021,Touresse2024}. 

Rotational surface motions are used to launch torsional Alfv\'{e}n waves into the open field of the northern polar coronal hole. The driving patch is centred on $(\theta,\phi)=(67^\circ,25^\circ)$ and has an angular width of $(\theta,\phi)=(2^\circ,4^\circ)$, corresponding at this latitude to an approximately circular region $\approx 60$\,Mm in diameter on the surface, Fig. \ref{fig:setup}(a). The radial magnetic field on the lower boundary within the patch is approximately uniform with $B_R\approx 20$\,G. The same expression for the surface motions is used as in \citet{Wyper2021}, but now the parameters are chosen so that the driving forms a circular annulus (similar to that used in \citet{Higginson2017}). This width was chosen to ensure sufficient resolution to resolve gradients in the driving profile across the driving region but the initial wave is still comparable in scale to those produced by the largest observed coronal jets \citep[e.g.][]{Shimojo1996,Savcheva2007}. 

The surface speed shown in Fig. \ref{fig:setup}(a) corresponds to Case 10 (see Table \ref{tab}) and peaks around $v_{max}=302$\,km\,s$^{-1}$ which is approximately $40\%$ of the Alfv\'{e}n speed on the surface ($\approx 750$\,km\,s$^{-1}$). We emphasise here that we are trying to reproduce torsional Alfv\'{e}n wave pulses entering the low corona from below, as would be produced by impulsive reconnection events or potentially fast, swirling chromospheric flows. An Alfv\'{e}n wave entering the volume from below is expected to have a rotational speed that is a large fraction of the Alfv\'{e}n speed, as opposed to the observed photospheric speeds on the order of $1$\,km\,s$^{-1}$ which we are not trying to reproduce. The time profile of the driving takes the form
\begin{equation}
F(t)=
\begin{cases}
\frac{1}{2}\left[1-\cos\left(\frac{2\pi t}{t_{pulse}}\right)\right] & 0\le t \le t_{pulse} \\
0 & t>t_{pulse}
\end{cases},
\label{eqn:driver}
\end{equation}
where $t$ is time and $t_{pulse}$ controls the duration of time that the driving pulse is applied for. We consider three sets of simulations, with fixed driving speeds and varied driving times summarized in Table \ref{tab}. The driving time is chosen so that across the three groups a comparable number of turns are injected into the wave, but over differing time frames (discussed further in \S \ref{sec:para}). As we shall see, when grouped in this manner the simulations display a remarkable similarity.

For simplicity in this initial investigation the driver is positioned in the northern polar coronal hole away from the pseudostreamer and helmet streamer so as to avoid open field regions with strong field line curvatures and cross field gradients in flow and Alfv\'{e}n speed. Future work will consider how these effects alter the results presented here. The edge of the driving profile defines a flux tube that extends into the coronal hole, and within which the wave is initially contained. Thus, the waves in each simulation have the same initial lateral extent ($\approx 60$\,Mm). A rough estimate of the initial radial extent is given by multiplying the full width at half maximum of the time profile ($t_{period}/2$) with the surface Alfvén speed ($\approx750$\,km\,s$^{-1}$) which ranges from $\approx 113$\,Mm ($0.15 R_\odot$) to $\approx 1300$\,Mm ($2.5 R_\odot$) across all the simulations, Table \ref{tab}. Figure \ref{fig:setup}(b) shows the cross section of the flux tube along which waves are launched. Field lines traced from the edge of the driving region are shown in pink. A field line traced from the centre of the driving patch/flux tube is shown in green. The solar wind solution has developed self consistently as a result of the relaxation process which formed the initial condition, and is purely a result of the Parker wind constrained by the flux tube geometry. The shading in Fig. \ref{fig:setup}(b) shows the Alfv\'{e}n speed in a plane of constant $\phi$, highlighting for example cross field gradients near the helmet streamer at the bottom of the panel. The dark grey contour shows where the plasma-$\beta = 1$ denoting where the wind moves from low-$\beta$ to $\beta \sim \mathcal{O}(1)$ in the outer volume.

Fig. \ref{fig:setup}(c) shows quantities measured along the green field line at the centre of the flux tube. The wind speed (blue) transitions to super-Alfv\'{e}nic ($v_{sw}=v_a$, crossing of red and blue curves) around $~12 \,R_\odot$, close to where $\beta = 1$ (yellow), around $~16\,R_\odot$. The propagation speed of an idealised Alfvén wave ($v_{sw}+v_a$) is shown in purple. Aside from a small bump near the surface, the propagation speed decreases rapidly until around $~12\,R_\odot$ after which it plateaus. Note that here (and throughout the paper) we use the unperturbed simulation velocity at $t=0$ for the background solar wind speed, so that $v_{sw}=|\mathbf{v}(\mathbf{x},0)|$. Similarly, unless otherwise specified $v_a = v_a(\mathbf{x},0)$, $\mathbf{B}_0=\mathbf{B}(\mathbf{x},0)$ and $\rho_0=\rho(\mathbf{x},0)$.

Shown in green is the expansion factor of the field line as a function of height. This is defined as
\begin{equation}
\label{eqn:exp}
f(R) = \left(\frac{R_{ph}}{R}\right)^2 \frac{B_{ph}}{B(R)},
\end{equation}
where $B_{ph}$ is the total magnetic field strength at the lower boundary. Typically, the expansion factor is calculated between two fixed radii, usually taken to be the photosphere and the source surface at $2.5 \,R_\odot$. The definition above generalises this so that $f(R)$ is the expansion factor between the lower boundary and the radius $R$ \citep[e.g.][]{Kohl2006}. Note that we use the definition with the total field strength \citep[e.g.][]{Wallace2020} rather than the just the radial component as is sometimes used \citep[e.g.][]{Riley2015,Dakeyo2024} as the waves propagate along the magnetic field direction which is not always radial in the volume. $f(R)$ grows rapidly and also plateaus around $12 R_\odot$, beyond which it remains near constant. The rapid growth indicates a super-radial expansion below $10\,R_\odot$, reducing to a radial expansion beyond $10\,R_\odot$, i.e. beyond this radius the field is expanding with a $|\mathbf{B}| \sim 1/R^2$ drop in field strength. We emphasise here that a total expansion along the flux tube of $ \mathcal{O}(10)$ is similar to values obtained from source surface extrapolations \citep[e.g.][]{Wallace2020,Dakeyo2024} and much higher than previous studies 
with spherical wedge domains where a radial field is employed \citep[e.g.][]{Karpen2017,Shoda2021,Touresse2024}. In fact, expansion factors of order $10$ are typical of quiet sun regions as would be observed away from the boundaries of coronal holes, but we note that expansion factors as high as several hundred have been found on the edge of active regions \citep[e.g.][]{Wallace2020}.

\begin{figure*}
\centering
\includegraphics[width=0.98\textwidth]{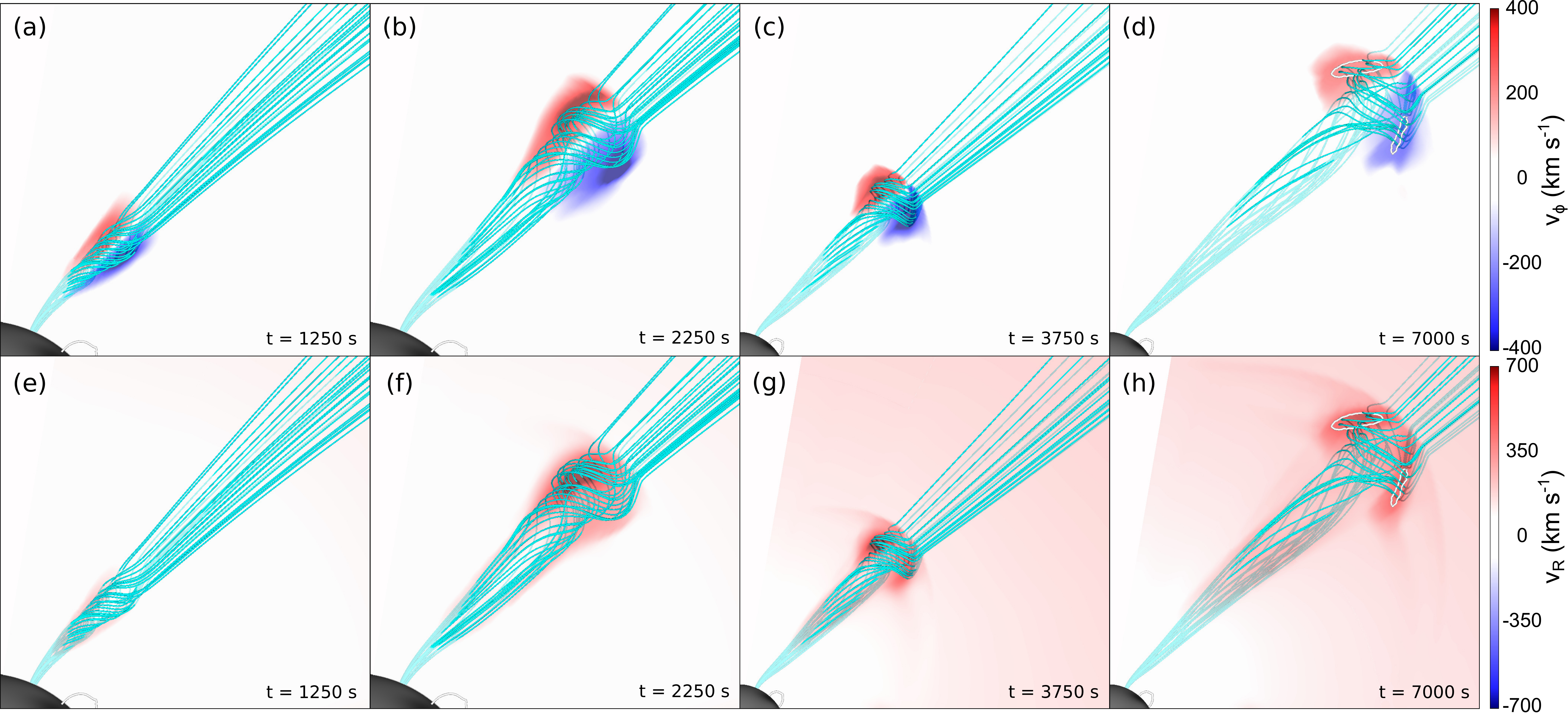}
\caption{Torsional Alfv\'{e}n wave evolution and switchback formation. Top panels: shading shows $v_\phi$ in planes of constant $\phi$ showing the rotation of the wave. Field lines are traced from fixed foot points on the lower boundary. Bottom panels: $v_R$ in the same planes showing the development of a shock front and strong radial jets. White contours show where $B_R=0$. Panels on the right half of the figure, i.e. (c), (d), (g) and (h), are viewed from further away than those on the left half. The front edge of the wave pulse at times $t = 1250$\,s, $2250$\,s, $3750$\,s, and $7000$\,s is located at $\approx 2.5\,R_\odot$, $\approx 3.2\,R_\odot$, $\approx 5\,R_\odot$, and $\approx 8R_\odot$, respectively -- see Fig. \ref{fig:ht}. An animation of panels (a) to (d) of the figure is available, showing the evolution of wave during the interval $0 \leq t \leq 24,000$\,s ($6$\,hrs $40$\,min) at at time cadence of $250$\,s (the animation duration is 6 s).} 
\label{fig:pole_vel}
\end{figure*}

\begin{figure*}
\centering
\includegraphics[width=0.98\textwidth]{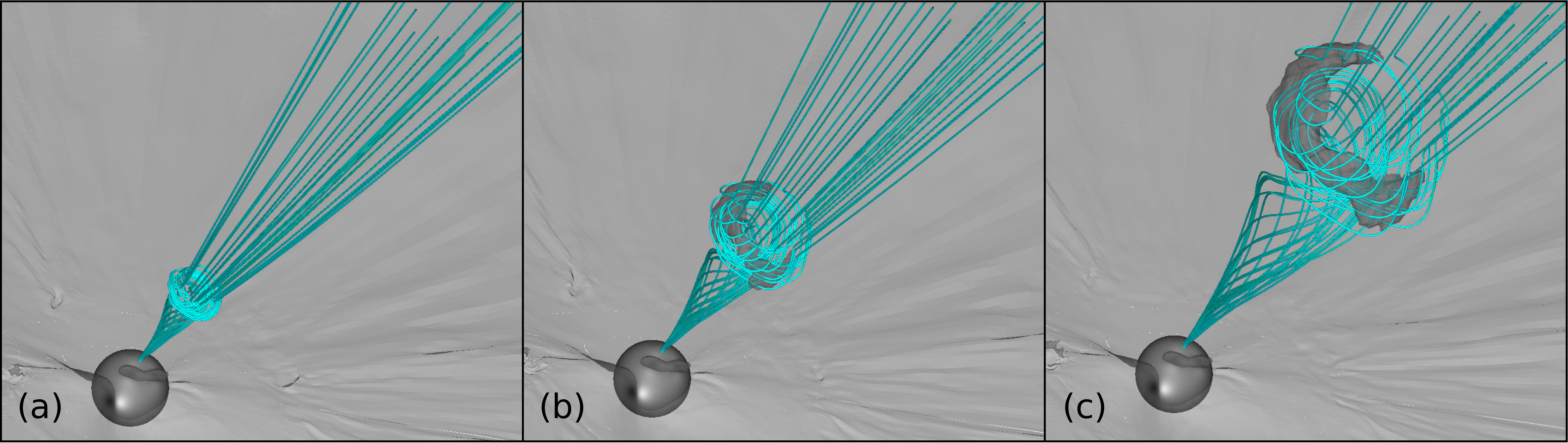}
\caption{3D rendering of the development of the switchback region. The grey iso-surface shows where $B_R=0$. An animation of the figure is available, showing the evolution of field lines passing through the switchback volume during the interval $0 \leq t \leq 24,000$\,s ($6$\,hrs $40$\,min) at at time cadence of $250$\,s (the animation duration is 6 s). } 
\label{fig:iso}
\end{figure*}

\section{Representative simulation example}
\label{sec:example}
\subsection{Overall evolution}
To discuss the general behaviour of torsional waves revealed by our simulations we now describe in detail a single representative case, before generalising our results in later sections. Here we focus on Case $10$ where $t_{pulse} = 1200$\,s ($20$\,min) with a maximum driving speed of $v_{max} \approx 302$\,km\,s$^{-1}$. Field lines within the wave as it is launched are shown in Fig. \ref{fig:setup}(a). We focus on this case as an example in particular as the time frame, speed and surface width of the wave are all consistent with a large coronal jet \citep[e.g.][]{Savcheva2007,Kumar2018}. In this case the driving injects just over one turn of twist into the wave pulse which is again typical of coronal jets \citep[e.g.][]{Liu2019}. 

The general evolution of the wave is shown in Fig. \ref{fig:pole_vel}, where the shading shows the out-of-plane and radial velocities and the white contours shows where $B_R=0$. As the torsional wave propagates upwards into the expanding flux tube, the reduction in the wave propagation speed ($v_{sw}+v_a$) causes the wave pulse to shorten as the back of the pulse catches up with the front \citep{Volk1973}, concentrating the rotating plasma, Fig. \ref{fig:pole_vel}(a)-(b). At the same time a strong radial flow component (of comparable magnitude to the out-of-plane velocity) forms within the main pulse Fig. \ref{fig:pole_vel}(f).

Around $5 R_\odot$ the aspect ratio of the rotating front region approaches $1$, Fig. \ref{fig:pole_vel}(c), with switchback reversals in $B_R$ following shortly after (the regions within the white contours), Fig. \ref{fig:pole_vel}(d). Throughout the evolution the rotational speeds slowly reduce as the wave propagates (compare panels (a) to (d)). The development of the strong radial flow component from initially modest values leads to a shock forming at the front of the wave around $t=3000$\,s (Fig. \ref{fig:pole_vel}(g)), which spreads into two apparent wings by $t=7000$\,s (Fig. \ref{fig:pole_vel}(h), discussed further below). The fastest radial jets reside within the regions where $B_R$ changes sign as is typical of switchback regions. Figure \ref{fig:iso} and the accompanying animation shows the 3D structure of the switchback volume (where $B_R<0$) showing it forms in a ring-like volume around the edge of the wave and that once the switchback forms it (somewhat surprisingly) maintains its overall structure until it exits the domain at $30 R_\odot$.

\begin{figure*}
\centering
\includegraphics[width=0.98\textwidth]{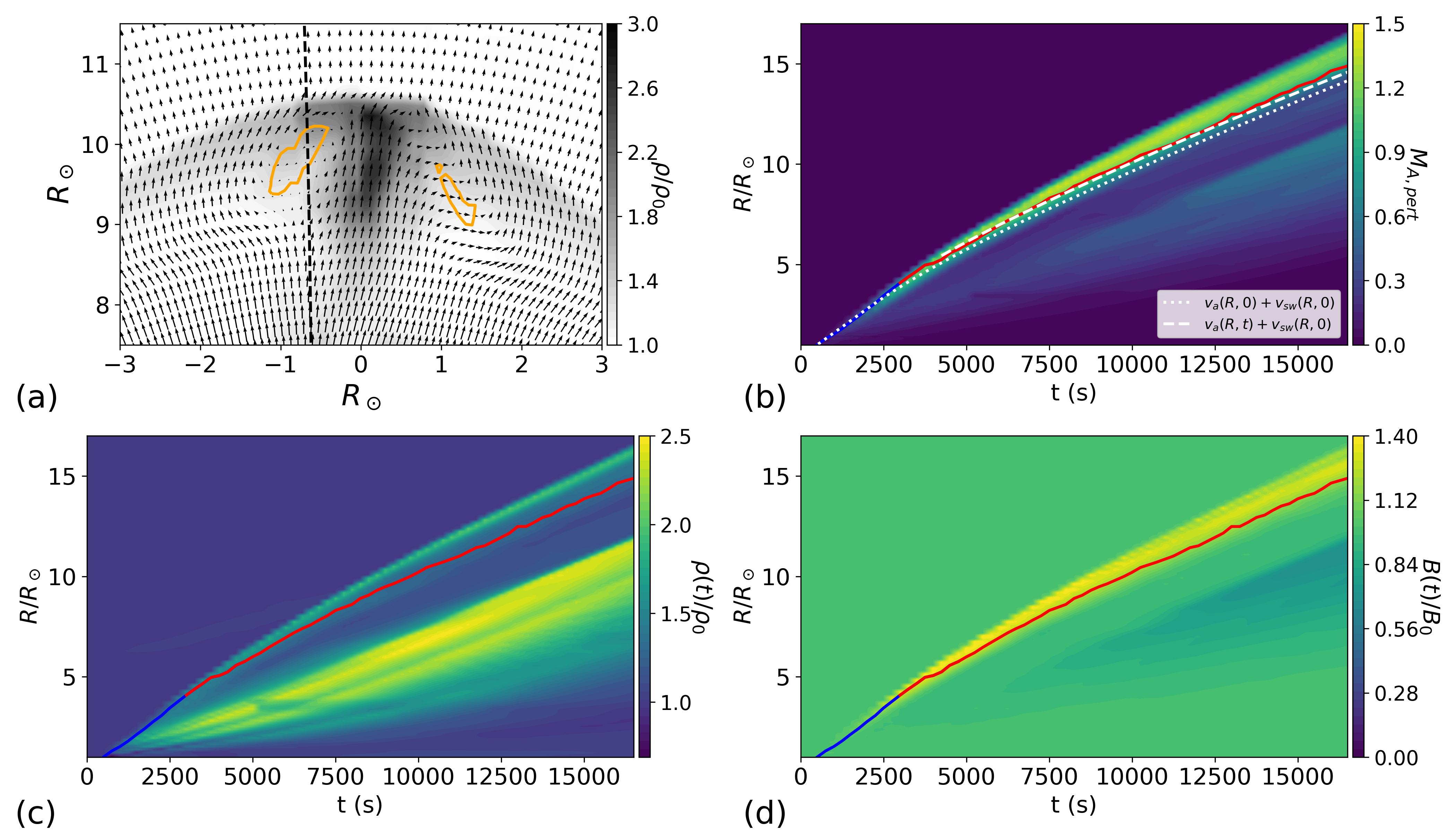}
\caption{{(a) Density enhancement in the vicinity of the switchback ring. Contours show $B_R=0$ in the plane ($\phi = 26^\circ$) while the arrows show the magnetic field direction. The dashed line shows the projection of the field line along which the samples for the height-time plots displayed in panels (b)-(d) were taken. In those three panels, the blue-to-red solid line indicates the radial coordinate of the position of maximum deflection ($R_{SB}$): Red indicates where $\theta_B > 90^\circ$, Blue where $\theta_B < 90^\circ$.
(b)} Height time plot of $M_{A,pert}$ along the green field line shown in Fig. \ref{fig:setup}(c). Dotted line: height-time curve from the background values of $v_{a}(R,0) + v_{sw}(R,0)=v_a(\mathbf{x}_{fl},0)+|\mathbf{v}(\mathbf{x}_{fl},0)|$, where $\mathbf{x}_{fl}(R)=(R,\theta(R),\phi(R))$ is the position along the unperturbed field line at $t=0$. Dashed line: height-time curve from the local {Alfv\'{e}n} speed combined with the background solar wind speed: {$v_a(R,t)+v_{sw}(R,0)$}. Note that this line starts at $t=4000$\,s. {(c) Height time plot of the density ratio,} $\rho(t)/\rho_0=\rho(\mathbf{x}_{fl},t)/\rho(\mathbf{x}_{fl},0)$. {(d) Height time plot of the magnetic field intensity ratio,} $B(t)/B_0=|\mathbf{B}(\mathbf{x}_{fl},t)|/|\mathbf{B}(\mathbf{x}_{fl},0)|$.
} 
\label{fig:ht}
\end{figure*}

\begin{figure}
\centering
\includegraphics[width=0.48\textwidth]{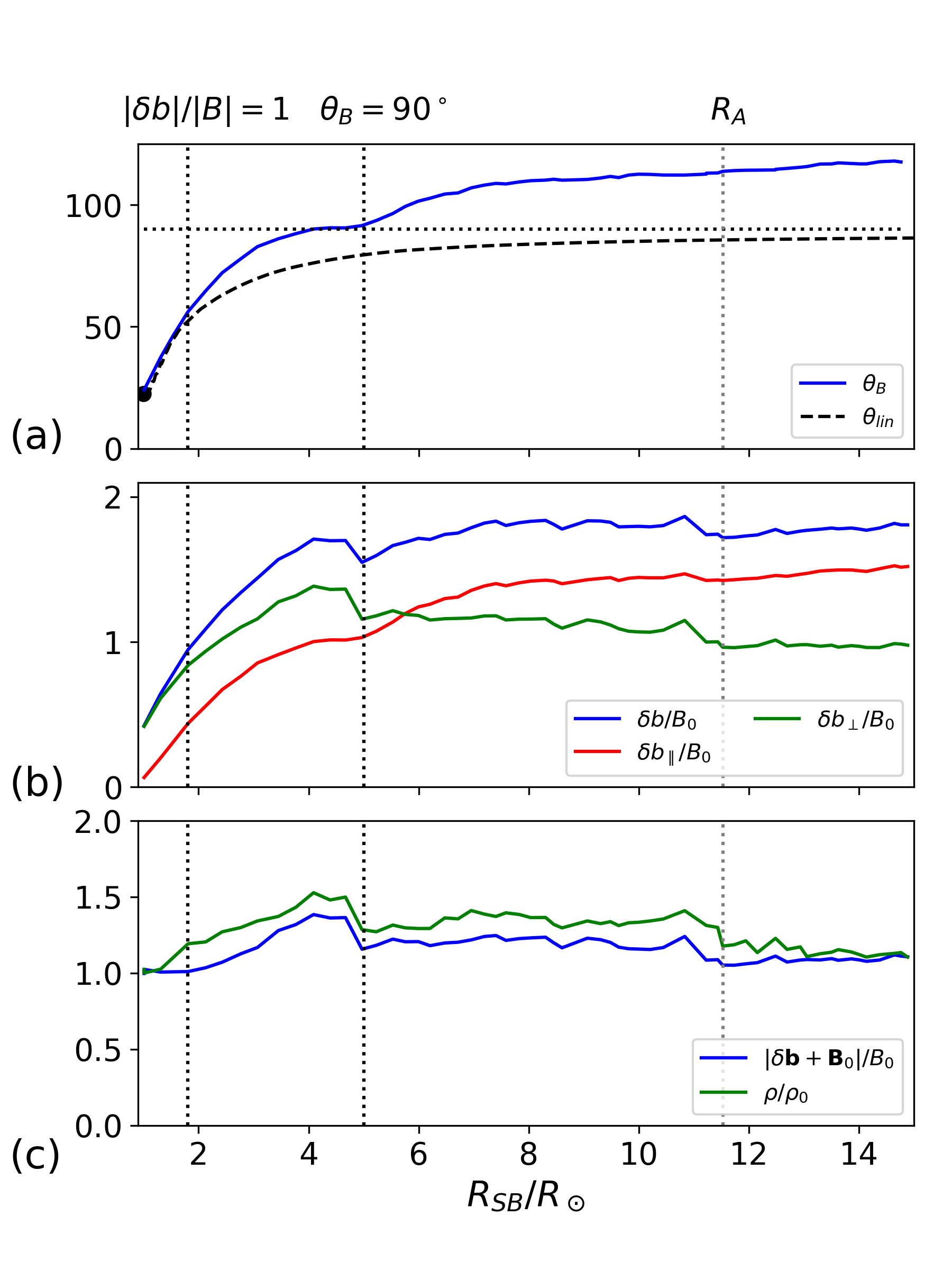}
\caption{Evolution of different quantities all measured at the location of the maximum deflection, $\theta_B$, as a function of the maximum deflection radial position, $R_{SB}$. (a) $\theta_B$ (blue line) and the estimate based on linear theory $\theta_{lin}$ (black dashed line, see text for details). The horizontal dotted line shows $\theta_B=90^\circ$. (b) Relative magnitude of the perturbation magnetic field ($\delta b=|\mathbf{B}(\mathbf{x},t)-\mathbf{B}_0|$ where $\mathbf{B}_0=\mathbf{B}(\mathbf{x},0)$ is the background magnetic field, blue line), its components parallel ($\delta b_\parallel = |\delta \mathbf{b}_\parallel|$, red line) and perpendicular ($\delta b_\perp = |\delta \mathbf{b}_\perp|$, green line) to $\mathbf{B}_0$. (c) level of spherical polarisation  ($|\delta \mathbf{b}+\mathbf{B}_0|/B_0$, blue line) and compressibility ($\rho/\rho_0$ where $\rho_0 = \rho(\mathbf{x},0)$, green line).} 
\label{fig:track}
\end{figure}

\begin{figure}
\centering
\includegraphics[width=0.48\textwidth]{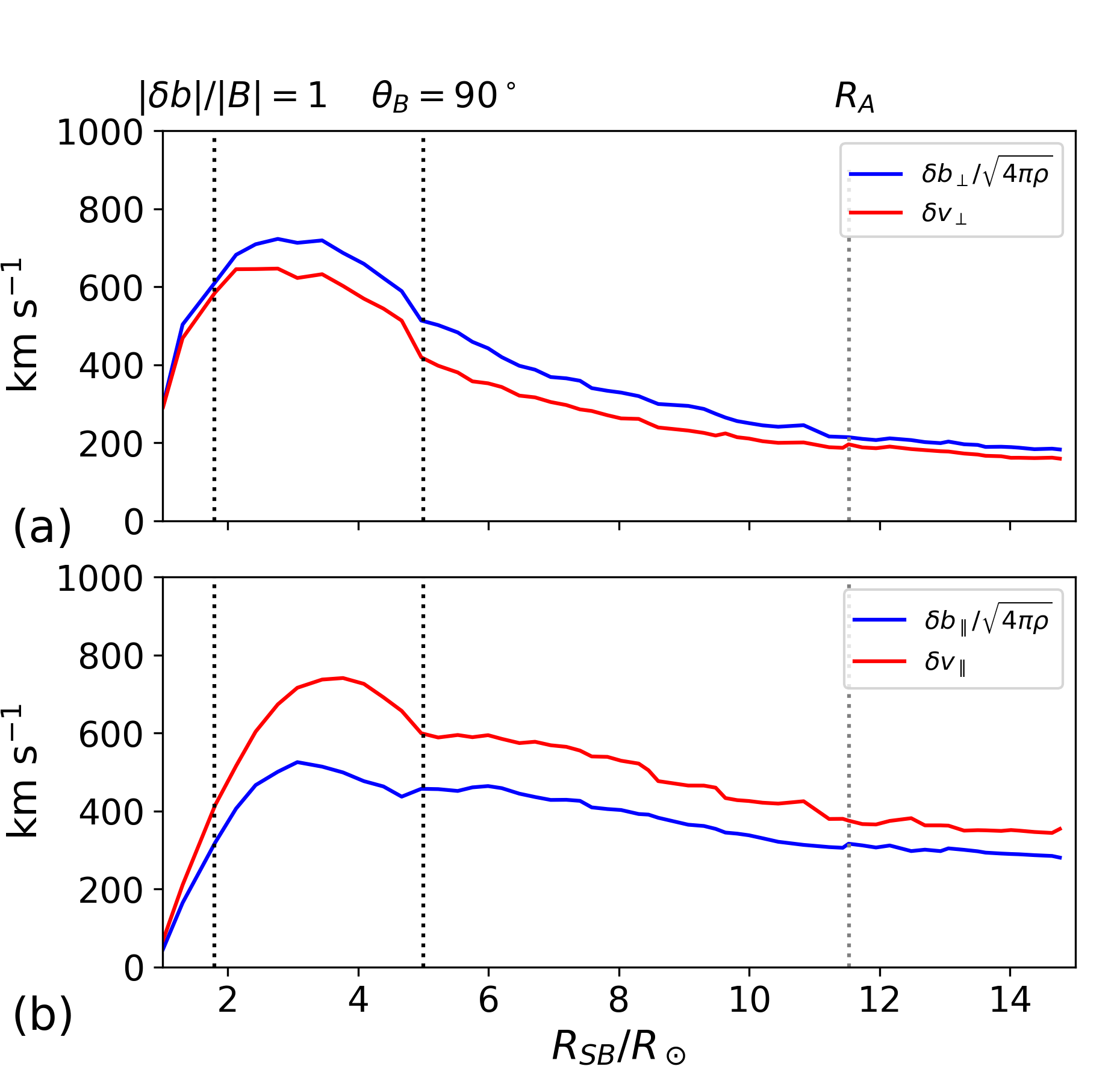}
\caption{Evolution of $\delta v_\perp$ vs $\delta b_\perp$ normalised to Alfv\'{e}n units (a), and $\delta v_{\parallel}$ vs $\delta b_{\parallel}$ normalised to Alfv\'{e}n units (b),  measured at the location of the maximum deflection, $\theta_B$, as a function of the maximum deflection radial position, $R_{SB}$. The vertical lines are the same as in Fig \ref{fig:track}} 
\label{fig:track2}
\end{figure}

\begin{figure*}
\centering
\includegraphics[width=0.98\textwidth]{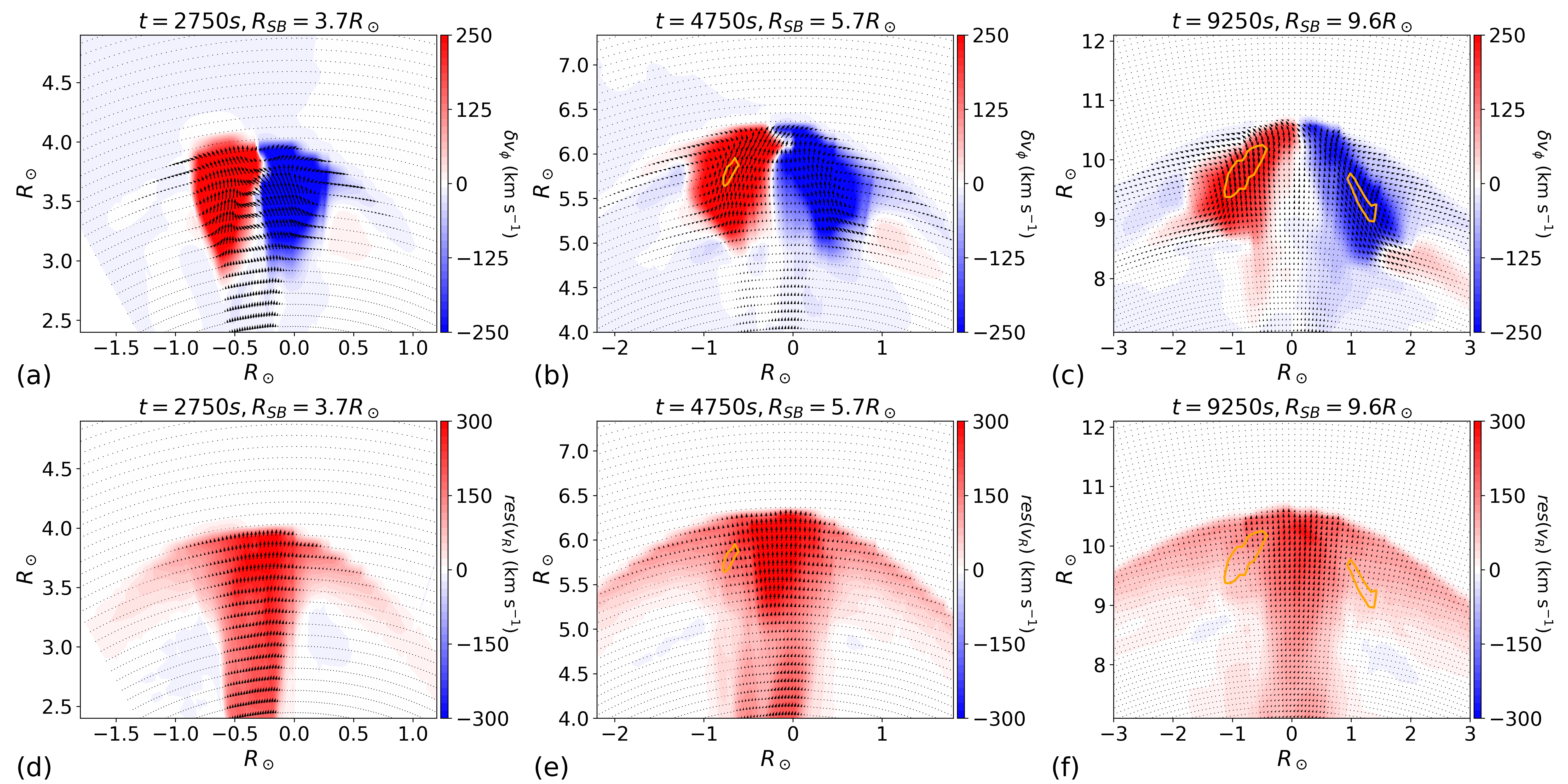}
\caption{The formation of the discs of inflow/outflow associated with the switchback in Case 10. (a)-(c): $\delta \mathbf{v} = \mathbf{v}(\mathbf{x},t)-\mathbf{v}(\mathbf{x},0)$ with arrows showing the in-plane motion, and the red-blue color map corresponding to the out-of-plane rotation. (d)-(f): the residual flow after subtracting the Alfv\'{e}nic component (Eqn. \ref{eqn:res}, see text for details). Arrows show the in-plane components, shading shows the magnitude of the radial component.} 
\label{fig:formation}
\end{figure*}

\begin{figure*}
\centering
\includegraphics[width=0.98\textwidth]{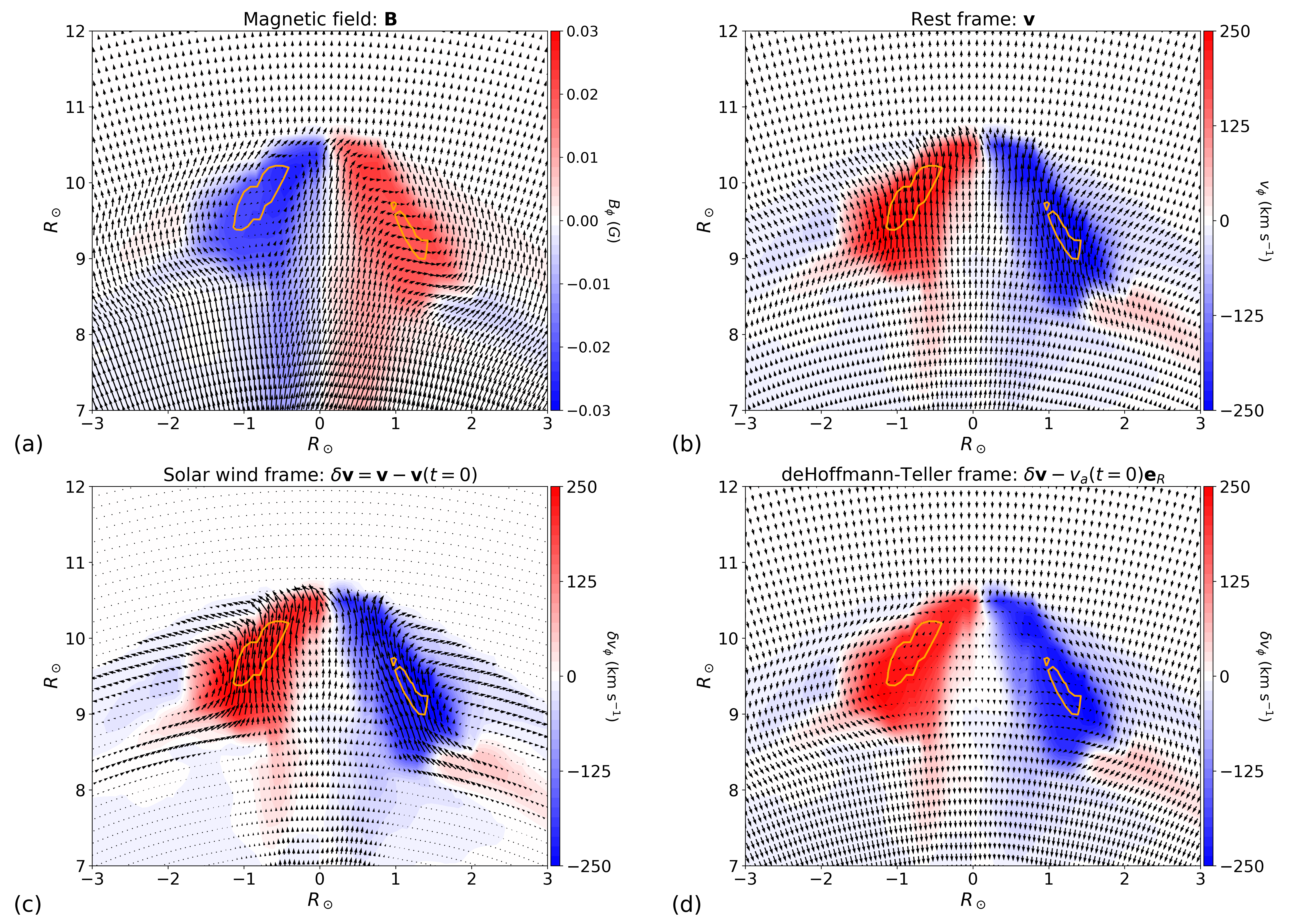}
\caption{Switchback structure at $t=9250$\,s ($R_{SB}=9.6R_\odot$) shown in different reference frames. (a) the magnetic field, (invariant of frame). (b) plasma velocity in the simulation rest frame. (c) plasma velocity in the solar wind frame. (d) plasma velocity moving approximately with the wave at $v_{sw}+v_a$ (the deHoffmann-Teller frame). Arrows show the in-plane vector field components while color-coded shading represents the out-of-plane vector field component.} 
\label{fig:frame}
\end{figure*}

\begin{figure*}
\centering
\includegraphics[width=0.98\textwidth]{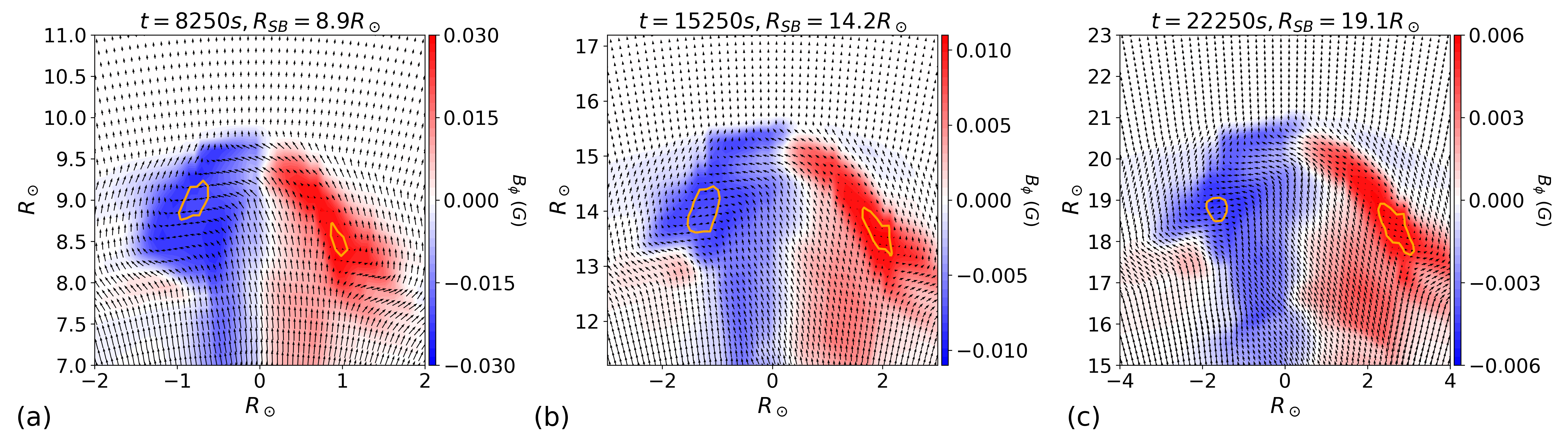}
\caption{The approximately self similar expansion of the switchback magnetic structure once formed. Arrows show the in-plane (constant $\phi$) vector magnetic field components while color-coded shading represents the out-of-plane magnetic field component ($B_\phi$). Note the increasing axis values from (a) to (c).} 
\label{fig:self}
\end{figure*}

\subsection{Switchback formation}
\subsubsection{Tracking the switchback position}
To better follow how the initial torsional Alfv\'{e}n wave evolves into the ring-like structure described above the local deflection angle relative to the unperturbed (background) magnetic field at $t=0$ was calculated everywhere within the volume, defined as
\[
\alpha(\mathbf{x},t) =\cos^{-1}{\left(\frac{\boldsymbol{B}\cdot\boldsymbol{B}_0}{|\boldsymbol{B}||\boldsymbol{B}_0|}\right)},
\]
where $\mathbf{B}_0 = \mathbf{B}(\mathbf{x},0)$. Since the computational mesh adaptively refines to follow the wave, the simulation grid differs at different times. Therefore, the grid was first uniformly sampled in the vicinity of the flux tube for all times. For each time the grid point with the maximal deflection was then identified, and its radial value: $R_{SB}(t)$, along with various other plasma parameters at this position were recorded. 

We define $\theta_B$ as the value of the deflection angle at $R_{SB}$, i.e. 
\begin{equation} \label{eq:defthetab}
\theta_B(R_{SB}(t)) = \text{max}\left[\alpha(\mathbf{x},t) \right],
\end{equation}
where we note that $\theta_B$ can be thought of as a function of $R_{SB}$ or equivalently time. Expressing $\theta_B$ as a function of radial distance ($R_{SB}$) in particular allows for a straightforward indication of the maximum deflection angle that could be measured at a given radial distance. We emphasise that this method follows the maximum deflection at each time, which does not necessarily correspond to the same position within the wave pulse itself. The position of greatest deflection is the product of both the wave motion and the wave evolution in the wave reference frame.

The red and blue solid line in Fig. \ref{fig:ht}(b) shows how $R_{SB}$ increases with time. The curve is blue when $\theta_B<90^\circ$ and red when $\theta_B>90^\circ$, i.e. once the switchback forms. For context the shading shows the Alfvén Mach number of the perturbation velocity in the solar wind frame ($M_{A,pert}$) calculated along the curve defined {by the dashed line shown in Fig. \ref{fig:ht}(a), which passes through the switchback volume.} This is defined by
\[
M_{A,pert}(R,t) = \frac{|\mathbf{v}(\mathbf{x}_{fl},t)|-|\mathbf{v}(\mathbf{x}_{fl},0)|}{v_a(\mathbf{x}_{fl},0)},
\]
where $\mathbf{x}_{fl}(R)=(R,\theta(R),\phi(R))$ is the position along the field line, $|\mathbf{v}(\mathbf{x}_{fl},0)|=v_{sw}(R,0)$ is the unperturbed initial solar wind velocity, and $v_a(\mathbf{x}_{fl},0)=v_a(R,0)$ is the initial Alfvén speed in the simulation volume. Note that this shows the speed of the perturbation in the solar wind frame relative to the Alfv\'{e}n speed, and differs from $M_A=v_{sw}/v_a$ which shows how fast the background solar wind is relative to the local Alfv\'{e}n speed.

The distribution of $M_{A,pert}$ shows that once the shock forms, the position of maximum deflection ($R_{SB}$) falls behind the shock. {It also shows that within the switchback the plasma is super-Alfv\'{e}nic ($M_{A,pert}>1$)}. The shock itself takes the form of a fast mode switch-on shock \citep[e.g.][]{Roberts2018,Fitzpatrick2022}, whereby ahead of the shock the plasma is unperturbed and a jump in $\delta b_\perp$ (increasing $|\mathbf{B}|$) {(Fig. \ref{fig:ht}(d))} and $\rho$ {(Fig. \ref{fig:ht}(a) and (c))} occurs across the shock. By its nature the increase in deflection angle across the shock cannot exceed $90^\circ$, and the shift of $R_{SB}$ to behind it reflects the fact that field lines within the main body of the wave continue to evolve to form deflections greater than $90^\circ$. {We also note that a density enhancement forms at the centre of the wave but that within the switchback the density is similar to background values (Fig. \ref{fig:ht}(a) and (c), discussed more below). A dense field-aligned flow is also excited that follows well behind the wave, Fig. \ref{fig:ht}(c), similar to flows noted in previous studies \citep[e.g.][]{Pariat2016,Karpen2017,Touresse2026}.}

The dotted white line shows the height-time curve corresponding to the propagation of a linear Alfvén wave front, calculated by numerically integrating $v_{sw}+v_a$ from an initial position on the lower boundary $R=R_\odot$ (see Appendix \ref{app:int} for details). The curve shows that in the early stages ($R_{SB}<5R_\odot$) the wave propagates at this speed, but deviates {slightly from this speed once the wave becomes non-linear and compressive effects start to play a role as the shock forms. The dashed line shows the numerically calculated height-time curve of an Alfv\'{e}n wave using the local (and space \& time varying) Alfv\'{e}n speed in the solar wind frame (initiated close to $R_{SB}$ after the shock forms), see Appendix \ref{app:int} for details. This curve closely matches the height-time evolution of $R_{SB}$, indicating that, as expected from the discussion in $\S$\ref{sec:review}, once formed the switchback continues to move at the Alfv\'{e}n speed in the accelerating solar wind, even when locally the plasma velocity is super-Alfv\'{e}nic (see Fig. \ref{fig:ht}(b)).}

\subsubsection{Evolution of local conditions within the switchback}

Figure \ref{fig:track}(a) shows $\theta_B$ as a function of $R_{SB}$. We contrast this against the expected deflection based upon a purely linear evolution of a circularly polarised wave (black dashed curve). Specifically, it can be shown that the conservation of wave action in the short wavelength limit gives that
\begin{align}
\label{eqn:theta}
\theta_{lin}(R) &= \tan^{-1}\left( \tan(\theta_{B,ph}) \left(\frac{B_{ph}}{B(R)}\right)^{1/2}\frac{(v_{a,ph}v_a(R))^{1/2}}{v_{sw}(R)+v_a(R)}\right),
\end{align}
where $X_{ph}$ indicates the value of the variable $X$ at the photosphere/lower boundary. In the above we have also used that $v_{sw,ph}\approx 0$, full details of the derivation are given in Appendix \ref{app:theta}. This can be further expressed in terms of the geometric properties of the open field via Eqn. \ref{eqn:exp} to give
\begin{equation}
\label{eqn:theta2}
\theta_{lin}(R) = \tan^{-1}\left( \tan(\theta_{B,ph}) \frac{R}{R_{ph}} f(R)^{1/2}\frac{(v_{a,ph}v_a(R))^{1/2}}{v_{sw}(R)+v_a(R)}\right).
\end{equation}
This expression assumes that the Alfvén wave contains only a perpendicular perturbation, i.e. that $\delta b = \delta b_\perp$ with $\delta b_\parallel =0$, neglecting the non-linear effects of spherical polarisation. A non-zero $\delta b_\parallel$ would lead to higher deflection angles, so this estimate provides a lower bound for the expected deflection. Each factor inside the bracket increases with radial distance so that in the linear regime the wave deflection plateaus at $90^\circ$. 

To estimate the initial deflection angle, we note that over an infinitesimal time period ($\delta t$), the field lines connected to the driving patch have their foot points advected a distance $v_{drive}\delta t$. While the wave front propagating the response of the field line to this motion into the volume travels a vertical distance from the boundary of $v_{a,ph}\delta t$. The maximum deflection occurs where the driving is fastest, giving the initial peak deflection on the lower boundary as
\[
\theta_{B,ph}\approx \tan^{-1}\left(\frac{(v_{drive})_{max}}{v_{a,ph}}\right),
\]
which in this case where $(v_{drive})_{max}\approx 0.8v_{a,ph}$ (Table \ref{tab}) equates to $\theta_{B,ph}\approx 39^{\circ}$. This initial value is shown as a black circle in Fig. \ref{fig:track}(a).

The two curves in Fig. \ref{fig:track}(a) match closely initially, while the wave remains in the linear regime (below $\approx 1.8 R_\odot$). After this the measured deflection increases faster than the linear rate, briefly plateaus at $90^\circ$ between $4$ and $5 R_\odot$, despite the smooth evolution of the wave during this period, before increasing once more. This plateau represents the shift in the position of maximal deflection within the wave, from within the original core of the rotation (which remains around $90^\circ$) to nearer its edge (which catches up and then surpasses $90^\circ$, discussed below).  

Figure \ref{fig:track}(b) shows that $\delta b/B_0$ follows a similar growth profile. Consistent with the match to the linear theory for $\theta_B$ and the propagation speed, in the early evolution the majority of the deflection is in the form of $\delta b_\perp$. A $\delta b_\parallel$ component then increases as the wave grows in amplitude. When the deflection surpasses $90^\circ$ the perpendicular component drops and the parallel component continues to increase, which is broadly consistent with the deflection increasing under the constraint of spherical polarisation. Figure \ref{fig:track}(c) shows the degree of spherical polarisation ($|\delta \boldsymbol{b}+\mathbf{B}_0|/B_0$) and local plasma compression ($\rho/\rho_0$) at the point of maximal deflection throughout this evolution. While in the linear regime $|\delta \boldsymbol{b}+\mathbf{B}_0|/B_0=1$, before growing slightly while the switchback forms and then returning to near one once the deflection exceeds $90^\circ$. The density perturbation also follows a similar trend. This shows that where the switchback forms behind the shock front (and so where the position of maximum deflection resides) the local plasma is relatively similar to typical switchback conditions, i.e. has little density and $|\mathbf{B}|$ variation. 

How Alfvénic the conditions are at the position of maximum deflection is shown in Fig. \ref{fig:track2}(a) and (b) which shows the parallel and perpendicular components of the perturbation velocity field ($\delta \mathbf{v}=\mathbf{v}(\mathbf{x},t)-\mathbf{v}(\mathbf{x},0)$) compared against the same components of $\delta \mathbf{b}$ normalised to Alfv\'{e}n units. For a perfect, isolated Alfv\'{e}n wave the curves should be identical. This is broadly true of the perpendicular components, particularly below $1.8 R_\odot$ before the wave becomes strongly non-linear and $\delta b/B_0$ surpasses $1$. Most notably though is that the parallel flow velocity is consistently larger than for a pure wave, particularly around the time the switchback is forming ($\approx 2R_\odot$ to $5 R_\odot$).

\subsubsection{Impact of the parallel flow}

Figure \ref{fig:formation} shows this flow and the role it plays in feeding back on the wave and forming the switchback ring. The top panels show $\delta \mathbf{v}$ (which shows the flows in the solar wind reference frame) in vertical planes crossing close to the centre of the wave at various times in the wave evolution. The shading shows the out-of-plane flow component, while the arrows show the in-plane flow direction. Note that the chosen planes differ in each panel as the background flux tube is slightly tilted with respect to planes of constant $\phi$. The bottom panels show the residual flows left after the Alfv\'{e}nic components have been removed, given by 
\begin{align}
res(\mathbf{v}) 
&= \delta \mathbf{v}+\frac{\delta \mathbf{b}}{\sqrt{\mu_0\rho}}\\
&=\mathbf{v}(\mathbf{x},t)-\mathbf{v}(\mathbf{x},0)+\frac{\mathbf{B}(\mathbf{x},t)-\mathbf{B}(\mathbf{x},0)}{\sqrt{\mu_0\rho(\mathbf{x},t)}},
\label{eqn:res}
\end {align}
where the plus sign comes from the fact that the perturbation fields for a positive background field are anti-correlated (Eqn. \ref{eqn:delta}). This quantity represents the combination of any reflected Sunward propagating Alfvén wave, compressive flows such as those associated with the shock and any induced parallel flows. All to some extent are likely present, but the shock and parallel flows are most clearly highlighted. It should be emphasised that these flows are coupled to and feedback on the Alfv\'{e}nic wave motion and are not independent of it.

Fig. \ref{fig:formation}(a)-(c) show that as the wave propagates, the shock forms at the wave front that spreads in width. At the same time strengthening cylindrically symmetric flows are set up that enter into the base of the wave front, pass through the most deflected magnetic field region (which becomes the switchback ring, orange contours) and pass out again parallel to the shock. These flows are absent in the lower panels, indicating they are the mostly incompressible Alfv\'{e}nic motions associated with the switchback structure. Specifically, they are the flows necessary to support the upflows within the switchbacks itself (discussed further below). The bottom panels (Fig. \ref{fig:formation}(d)-(f)) show that the induced parallel (to the background field) flows pass up through the centre of the wave, and in particular the switchback ring forms on the shear layer between the inner flows and the surrounding (stationary in this frame) solar wind, Fig. \ref{fig:formation}(f). In a sense, the Alfv\'{e}n wave is riding the upward flows, but since the centre of the wave moves faster than the sides, this leads to the field lines connecting the inner and outer regions of the wave moving through a deflection of $90^\circ$ and becoming Sunward directed in order to maintain the connection between the two regions. This evolution is consistent with previous studies on the effect of background shear flows on Alfv\'{e}n waves \citep[e.g.][]{Toth2023}, although in our case the shear flows appear to be self generated rather than externally imposed.

This evolution is also similar in a sense to the vortex ring formation that occurs at the front of hydrodynamic flow jets, e.g. smoke rings and bubble rings, but with some key differences. In hydrodynamic vortex rings the flow recirculates as it wraps around the toroidal axis of the vortex ring. In our case there is no recirculation, rather a disc of near parallel inflows/outflows form. The key difference appears to be the desire of the whole structure to remain close to Alfv\'{e}nic (and therefore near incompressible) and also spherically polarised. Recall from \S \ref{sec:review} that in a near Alfv\'{e}nic evolution, $\delta \boldsymbol{b} \approx -\sqrt{\mu_0\rho}\delta \boldsymbol{v}$. A recirculation of the flow within the wings would imply a Sunward flow component, which gives rise to a positive increase in $\delta \boldsymbol{b}$. However, this would violate local spherical polarisation as this is the same direction as the background magnetic field (and so $|\delta \boldsymbol{b}+\boldsymbol{B}_0|> |\boldsymbol{B}_0|$). Thus, the inflow/outflow required to sustain the upflow within the structure are constrained to sit adjacent to one another rather than to join into a full rotation.

Importantly, despite the deviation from Alfv\'{e}nicity as the switchback forms, as shown in Fig. \ref{fig:track}(c) once the structure passes $10 R_\odot$ and enters into a region where Parker Solar Probe would encounter it, the structure has returned to being close to Alfv\'{e}nic and spherically polarised. This is further shown in Fig. \ref{fig:frame} which shows a cross section of the magnetic field and flow velocities in various frames of reference as it passes $10 R_\odot$. Note that as expected from the discussion in \S \ref{sec:review} the total flow velocity $\mathbf{v}$ in the deHoffmann-Teller frame (Fig. \ref{fig:frame}(d)) is close to anti-correlated with the total magnetic field $\boldsymbol{B}$ (Fig. \ref{fig:frame}(a)).

\subsection{Self-similar expansion}
Once the magnetic structure (with the switchback volume in which $\theta_B>90^\circ$ embedded) has formed the overall shape of the perturbation is maintained as it expands. In a sense, the action of the shear flow moves the wave from one slowly evolving local equilibrium (the steepening torsional Alfv\'{e}n wave) to another (the rotating MHD vortex ring/disc). Surprisingly, the new switchback structure does not then behave like the earlier Alfv\'{e}n wave. The expectation from solar wind wave theory \citep[e.g.][]{Volk1973} is that Alfv\'{e}nic perturbations have a parallel length scale that scales with the wave's propagation speed, i.e. $l_\parallel \propto (v_{sw}+v_a)$. While flux conservation arguments imply that the perpendicular length scale $l_\perp \propto B^{-1/2}$. Together this implies that the structure should become wider and flatter, i.e. it should ``pancake", as it propagates. 

Although this is true of the initial wave, once the switchback forms it expands in a roughly self-similar manner whereby the width and length of the magnetic structure increase in a comparable way. This is shown qualitatively for this simulation in Fig. \ref{fig:self} (and will be further quantified in \S\ref{sec:para}). Panels (a) to (c) show the magnetic field at three times once the structure forms. Despite expanding by around a factor of two between 9 solar radii, Fig. \ref{fig:self}(a), and 19 solar radii, Fig. \ref{fig:self}(c), the overall aspect ratio and structure is closely maintained. This is likely a result of the continued action of the parallel flow jet, with the ongoing (but weakening) shear balancing the tendency of the wave to flatten out.

\begin{figure*}
\centering
\includegraphics[width=0.98\textwidth]{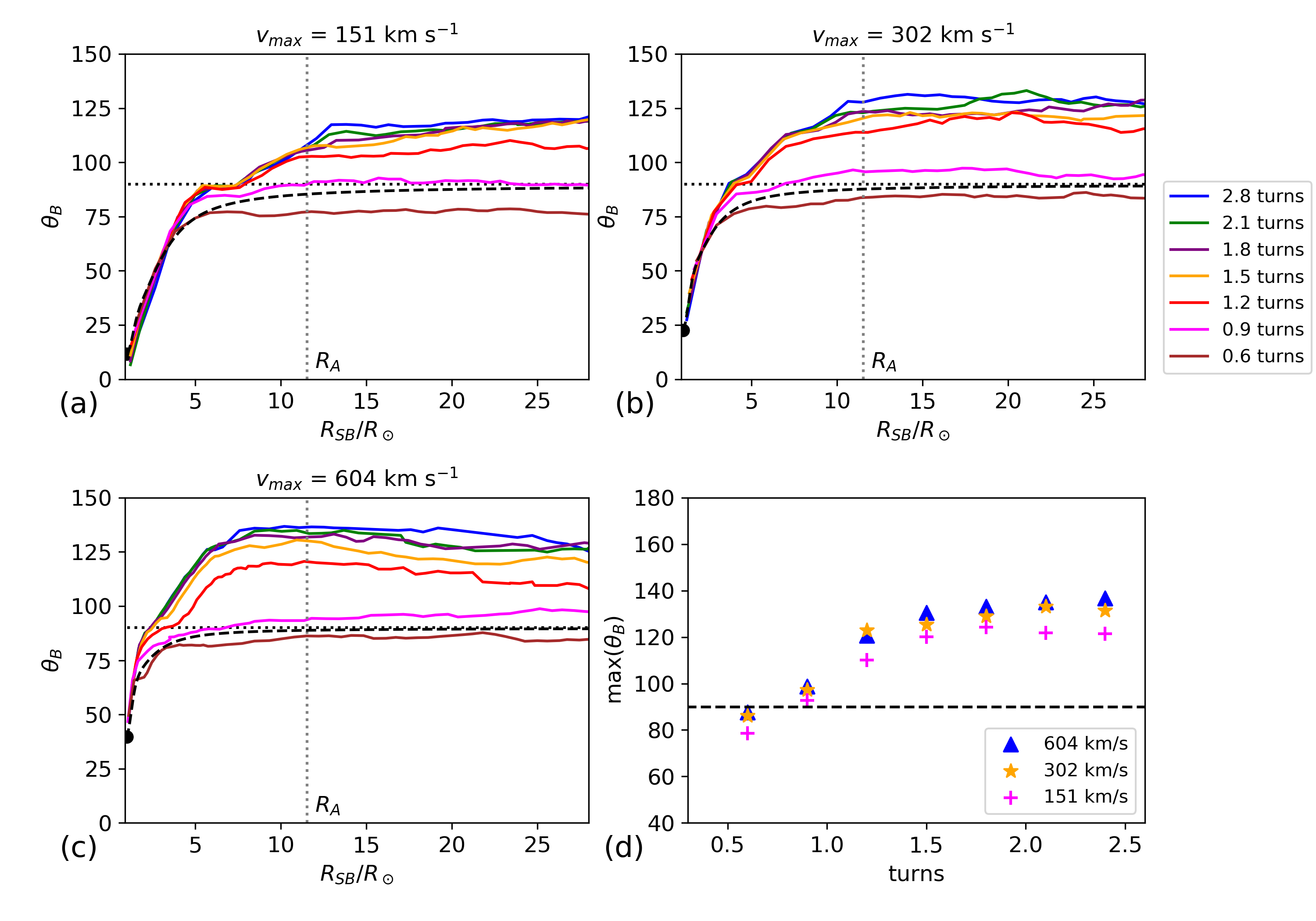}
\caption{Maximum magnetic deflection ($\theta_B$) vs its radial position ($R_{SB}$) in each simulation. (a) Cases 1 to 7 ($(v_{drive})_{max}=604$\,km\,s$^{-1}$), (b) Cases 8 to 17 ($(v_{drive})_{max}=302$\,km\,s$^{-1}$), (c) Cases 18 to 21 ($(v_{drive})_{max}=151$\,km\,s$^{-1}$). (d) the maximum field line deflection angle in the volume across all time outputs for each simulation compared to the maximum number of turns within the wave. Dashed black lines show the expected deflection from linear theory (Eqn. \ref{eqn:theta}). Horizontal dotted lines show $\theta_B=90^\circ$.} 
\label{fig:theta}
\end{figure*}

\begin{figure*}
\centering
\includegraphics[width=0.98\textwidth]{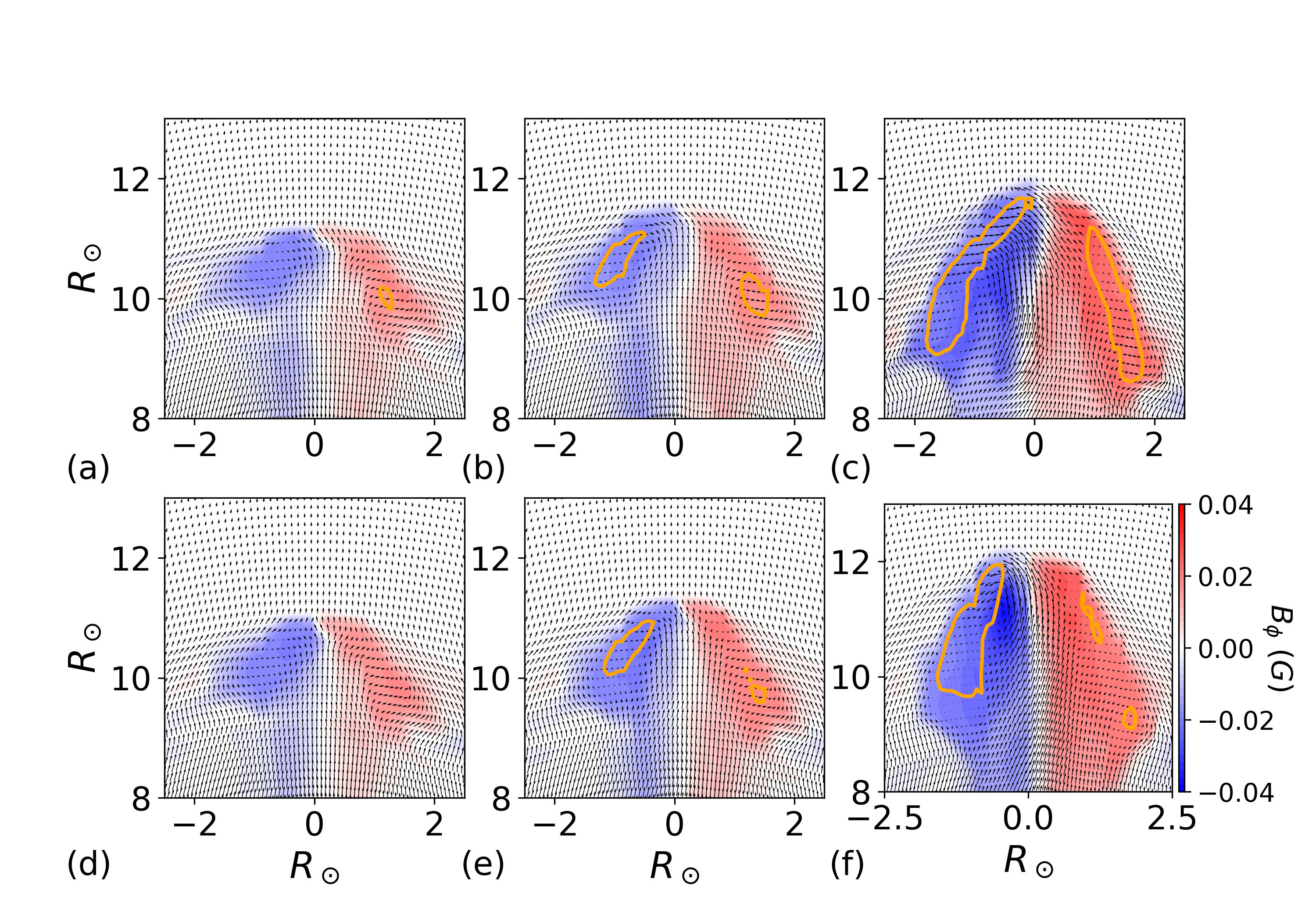}
\caption{The switchback structure at comparative radial distances in different simulations. Left column: peak turns $\approx 0.9$. Middle column: $\approx 1.2$. Right column $\approx 2.4$. Top row: peak driving speed $(v_{drive})_{max}= 604$\,km\,s$^{-1}$. Bottom row: peak driving speed $(v_{drive})_{max}= 302$\,km\,s$^{-1}$. Orange contour: $\theta_B=90$.} 
\label{fig:slice}
\end{figure*}

\begin{figure}
\centering
\includegraphics[width=0.48\textwidth]{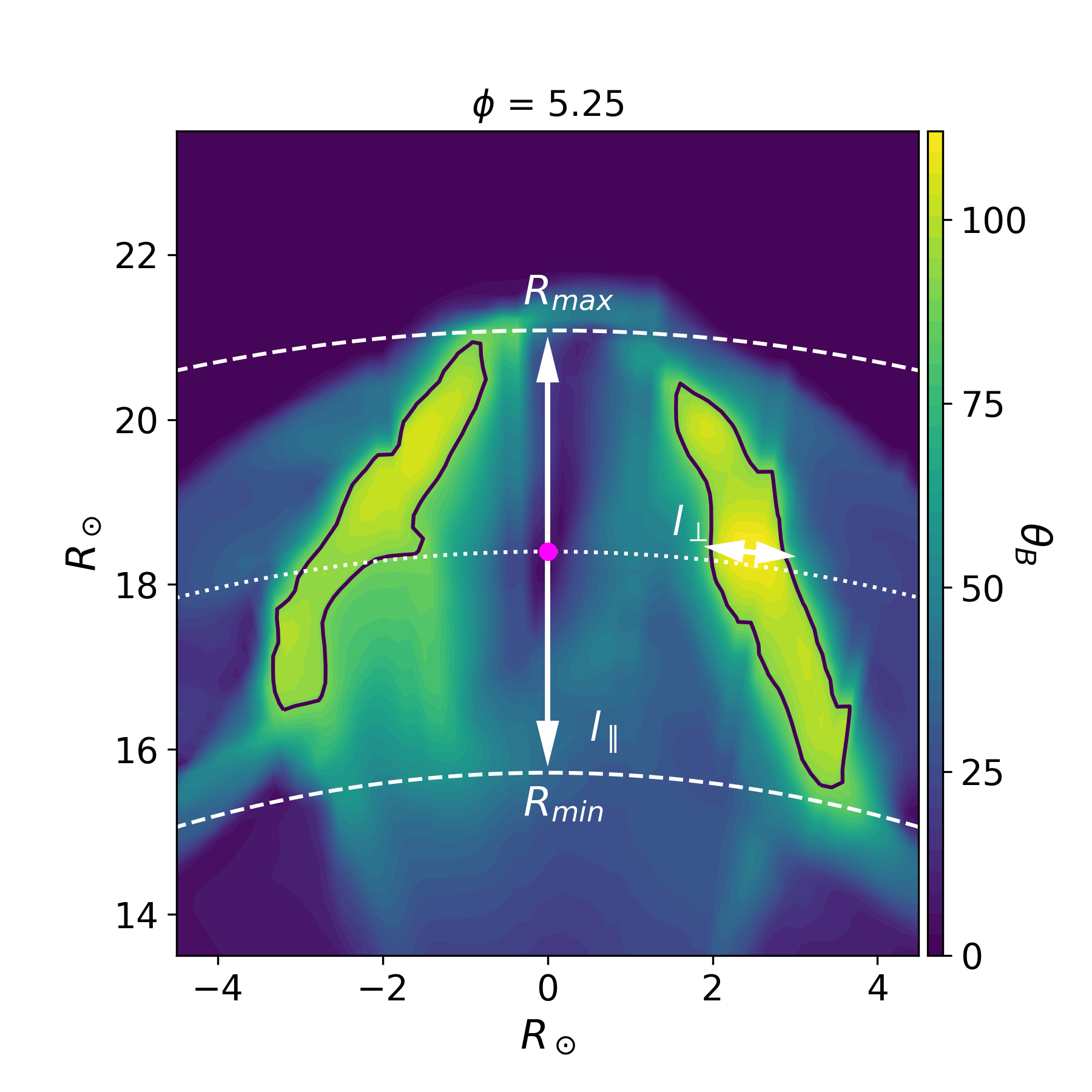}
\caption{Illustration of how $l_\parallel$ and $l_\perp$ are measured (Case 7). The black contours shows $\theta_B=90^\circ$. $l_\perp$ is taken as an average on the surface shown by the dotted line.} 
\label{fig:len_dig}
\end{figure}

\begin{figure*}
\centering
\includegraphics[width=0.98\textwidth]{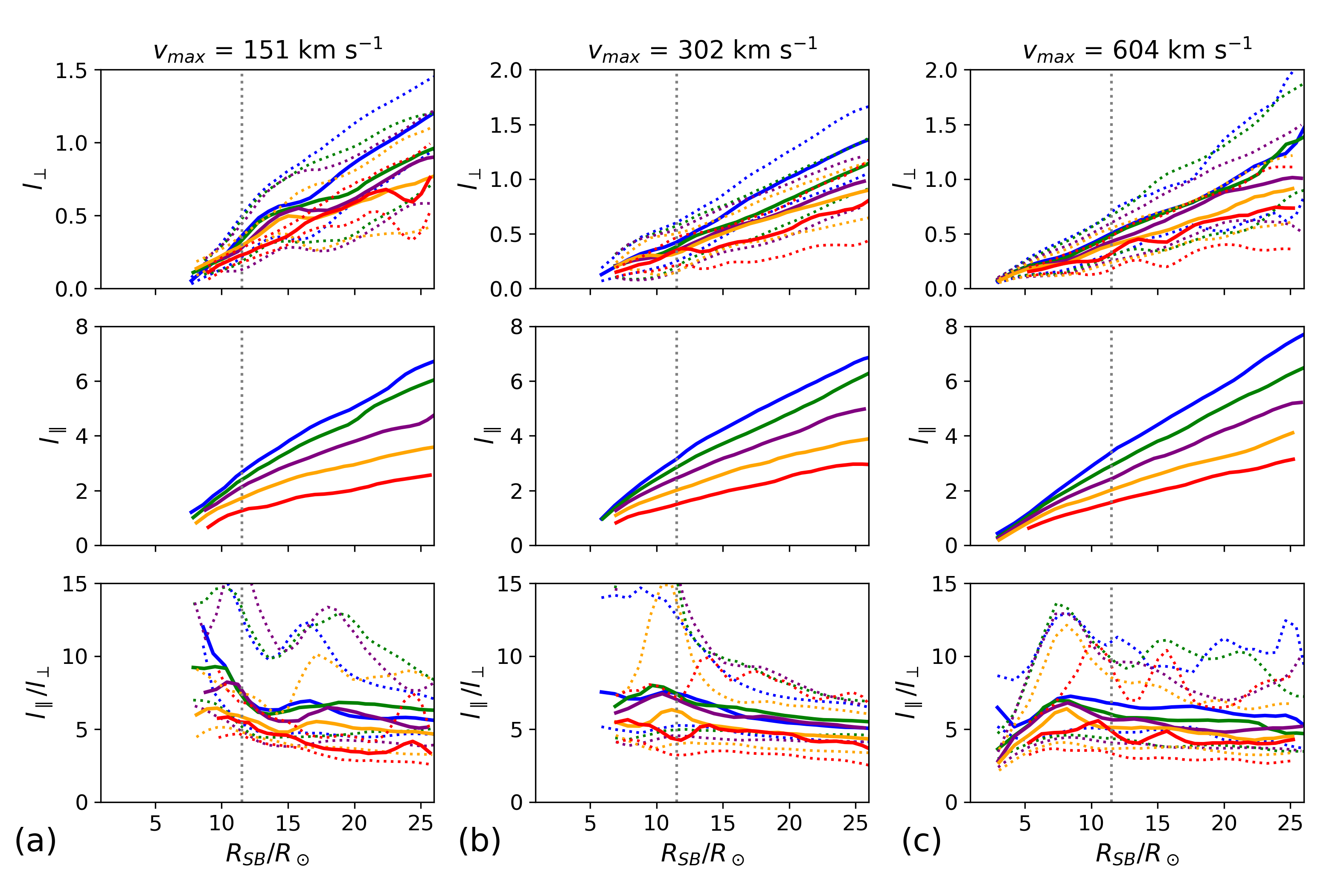}
\caption{Switchback length scales as a function of solar radius. Colour scale as in Fig. \ref{fig:theta}. Note Cases with $0.6$ and $0.9$ turns have been omitted as they have a small or non-existent volume where $\theta_B>90^\circ$. The dotted {vertical} line shows the Alfvén radius $R_A$ (where $v_{sw}=v_a$). {Dotted colored lines show the standard deviation in the $l_\perp$ and $l_\parallel/l_\perp$ measurements, with the color matching the respective mean value curve}.} 
\label{fig:multi_len}
\end{figure*}

\begin{figure}
\centering
\includegraphics[width=0.48\textwidth]{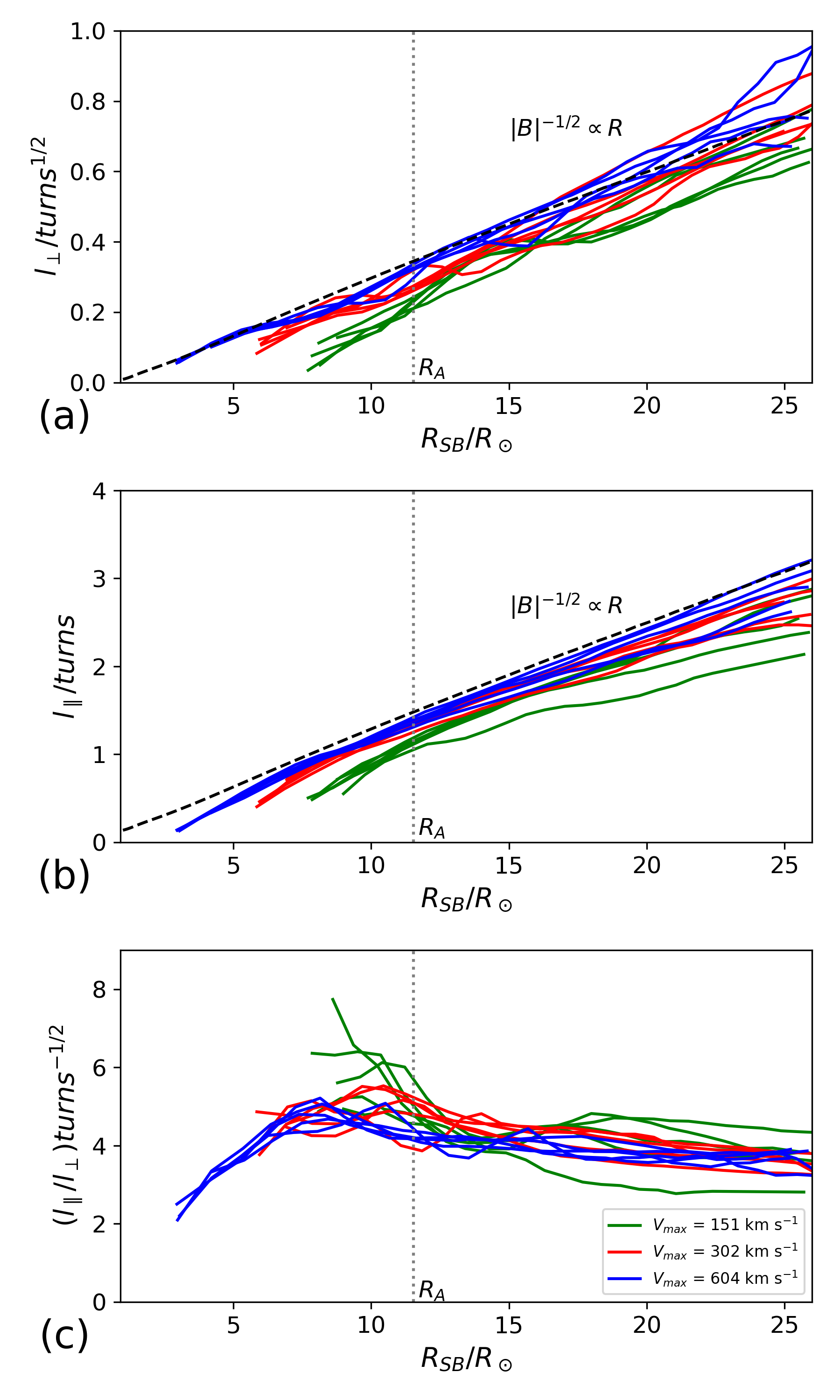}
\caption{Switchback length scales as a function of switchback radial position (in solar radii). (a) normalised $l_\parallel$, (b) normalised $l_\perp$, (c) normalised aspect ratio.} 
\label{fig:comb_len}
\end{figure}

\begin{figure*}
\centering
\includegraphics[width=0.98\textwidth]{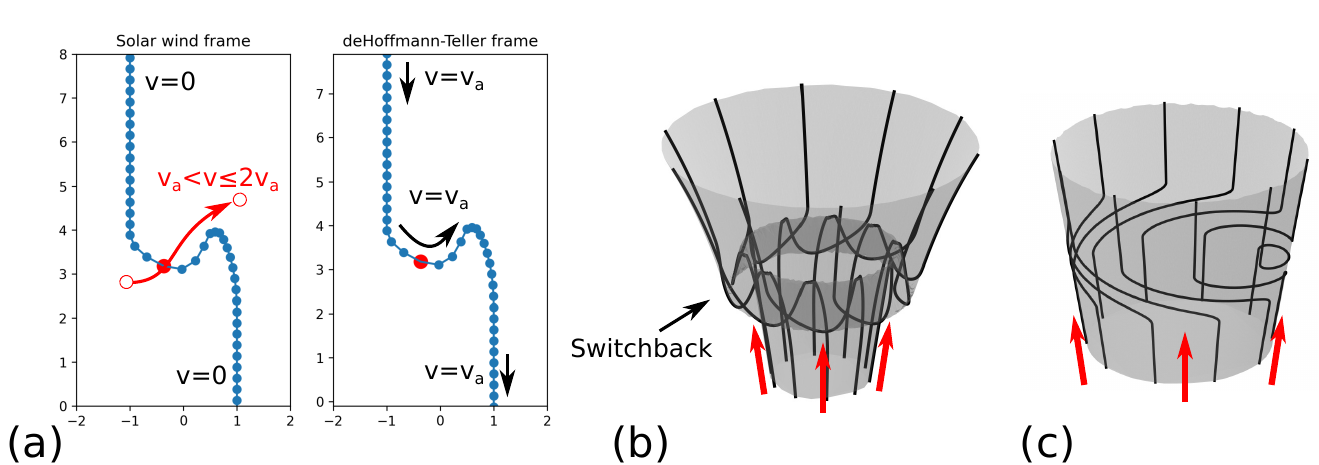}
\caption{(a) Flows in an idealised switchback in the solar wind frame and deHomann-Teller frame (moving with the wave). Fluid elements are shown in blue and red. (b) schematic of the flux surface and field lines produced by the wave induced flows. The switchback ring forms on the underside of the fold. (c) diagram showing how switchback formation (without topology change) is inhibited in a uniformly expanding field under the action of field aligned flows due to the axisymmetry. An animation of panel (a) of the figure is available, showing fluid elements moving in the idealised model in the two frames of reference with time in arbitrary units (the animation duration is 4 s).
} 
\label{fig:schem}
\end{figure*}

\begin{figure*}
\centering
\includegraphics[width=0.9\textwidth]{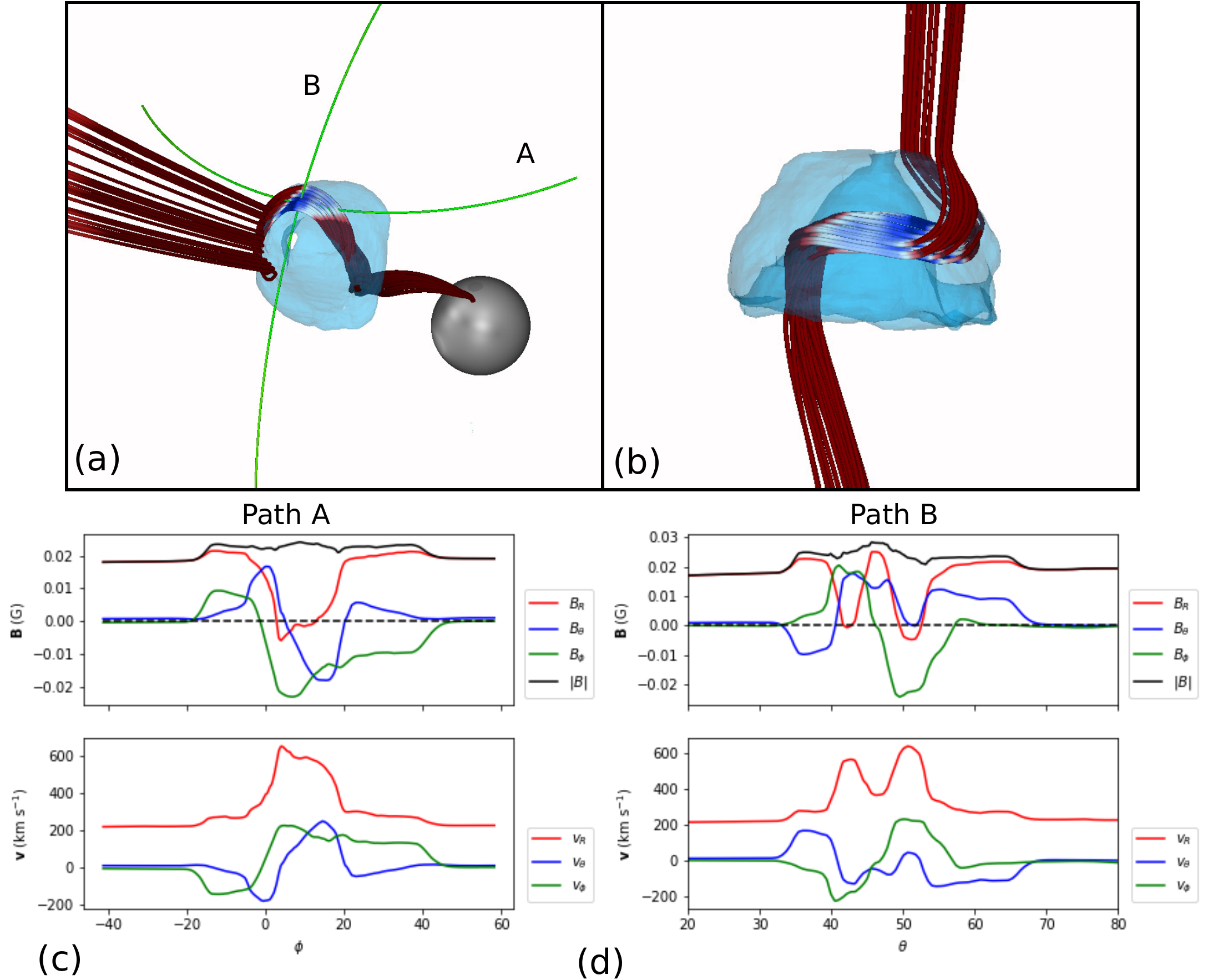}
\caption{(a) and (b) sample field lines within the switchback volume coloured by $B_R$ (max/min values: $\pm 0.005$G). Blue isosurface: $|\mathbf{v}|=500$\,km/s. (c) and (d) magnetic and velocity components measured along trajectories shown in green in (a).} 
\label{fig:probes}
\end{figure*}

\begin{figure*}
\centering
\includegraphics[width=0.9\textwidth]{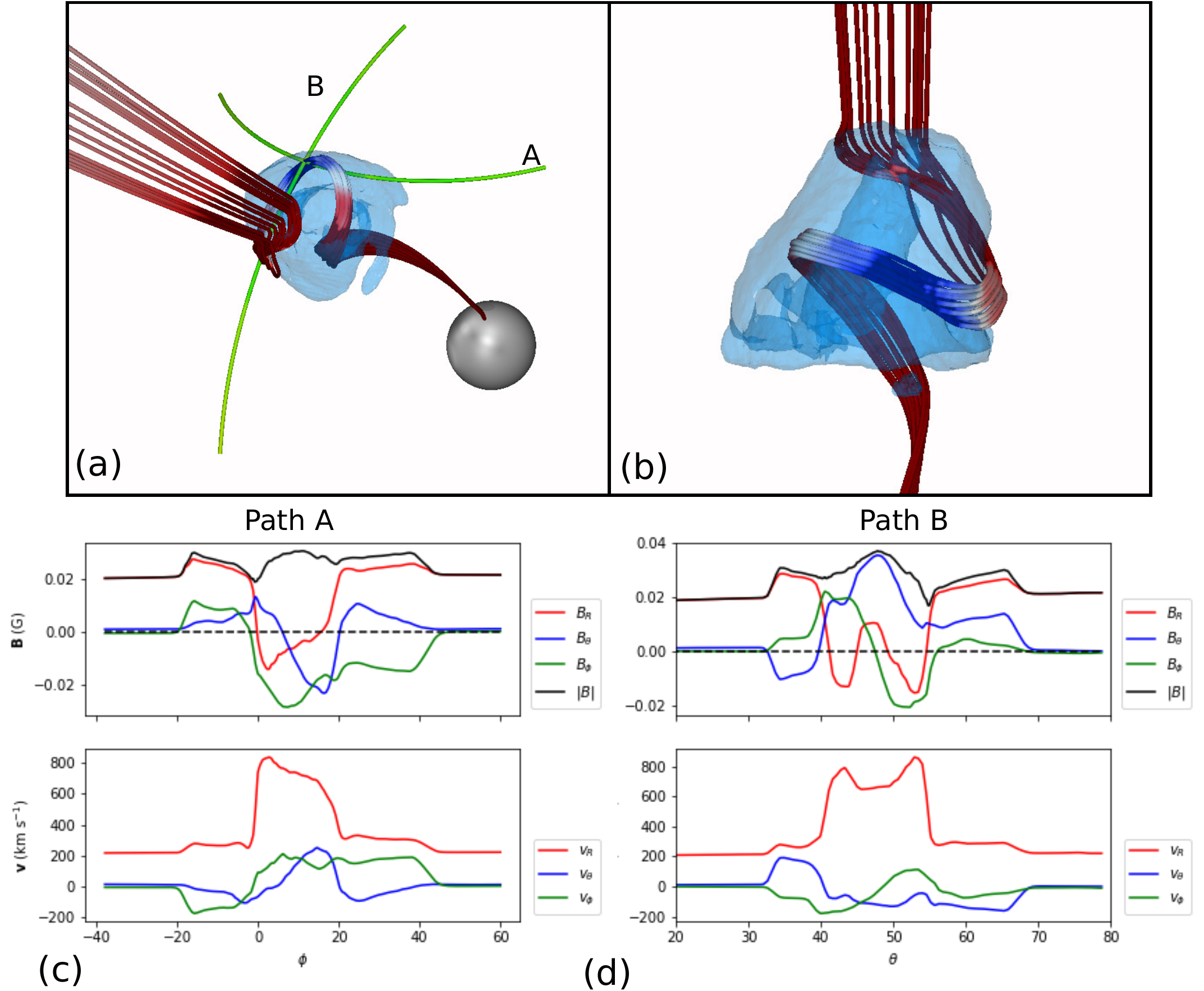}
\caption{(a) and (b) sample field lines within the switchback volume coloured by $B_R$ (max/min values: $\pm 0.005$G). Blue isosurface: $|\mathbf{v}|=600$\,km/s. (c) and (d) magnetic and velocity components measured along trajectories shown in green in (a).} 
\label{fig:probes2}
\end{figure*}

\section{Parameter dependence}
\label{sec:para}
\subsection{Magnetic deflection}
Having established the overall stages of evolution, we now consider how sensitive the ultimate switchback is to the properties of the initial wave. Specifically, we now vary the driving time and surface driving speed to vary the timeframe over which the same twist is injected into the wave. Our motivation for varying this is that coronal jets, jet-lets and surface swirling motions have widely varying durations, initial deflections and energies. Although our parameter ranges are not fully encompassing of all torsional waves associated with these phenomena, this initial study does show some remarkable trends which we will argue can be applied more generally. 

Table \ref{tab} lists the three groups of simulations and the maximum number of turns injected by the driving in each. The peak number of turns scales as 
\[
turns \approx \frac{t_{pulse}}{2}\left(\frac{v_\perp}{r}\right)_{max},
\]
where $r$ is the surface radial distance from the center of the driving rotation, so that $(v_\perp/r)_{max}$ is the maximum angular rotation (in both space and time) within the driving region. The factor of $t_{pulse}/2$ arises from integrating the driving time profile (Eqn. \ref{eqn:driver}) in time, i.e. $\int_{0}^{t_{pulse}}F(t)\,dt$. Therefore, the peak number of turns depends linearly on driving speed and time, so by varying them inversely the same number of turns are injected. The peak number of turns for Case 10 were measured by eye with the rest extrapolated using the above relation. Test field line tracing in the other cases confirms the peak number of turns are accurate within we estimate an $\approx 10\%$ error. Each set of simulations is grouped by their number of turns, varying from $\approx 0.6$ turns to $\approx 2.4$ for each driving speed. 

Figure \ref{fig:theta} summarizes the maximum deflection angle ($\theta_B$, cf. Eq. \ref{eq:defthetab}) as a function of radial distance in the three groups of simulations. The plot lines are coloured according to the peak number of turns injected, with for example all red curves injecting a maximum of $\approx 1.2$ turns into the wave over progressively faster time frames. Also plotted for context is the maximum deflection based on linear theory ($\theta_{lin}$), black dashed line. What these plots show is that when the twist is injected more slowly (panel (a)), so that the twist is spread out over a longer initial wave pulse and has a smaller initial maximal field line deflection, the field reversals develop further out ({vs} panels (b) and (c)). That is to say they take longer to develop as the wave propagates (compare where the curves pass $\theta_B \approx 60^\circ$ in each panel, beyond which they enter the non-linear regime). This is to be expected, and simply represents the additional travel distance required for the wave to steepen in its early evolution.  

The least twisted cases with $\approx 0.6$ turns (brown curves) remain essentially within the linear regime for their entire evolution, and actually drop below the linear estimate further out (likely as a result of dissipation). However, for the rest of the simulations once each wave reaches the non-linear phase the ultimate deflection is very similar for a given amount of twist injected. Interestingly, although the peak deflection in each generally increases when more twist is initially present in the wave, the effect saturates after around $1.2$ turns, with the peak deflection plateauing around $130^\circ$. This is shown in Fig. \ref{fig:theta}(b), which plots the maximum deflection recorded for each simulation vs maximum turns injected. The reason for the plateau is unclear but may be linked to the lengthening of the switchback region along the parallel direction, discussed more below.  

Another key finding is that since the magnetic structure evolves self-similarly once the switchback forms, when the switchback is viewed further out the ultimate magnetic structure is remarkably similar when the same amount of twist is injected. This is shown in Fig. \ref{fig:slice} which shows the magnetic structure of three sets of switchbacks as they pass $10 R_\odot$ where they would be encountered by PSP. With increasing twist the switchback ring lengthens along the radial direction (left to right across the panels), but more importantly for a given amount of twist, the ultimate switchback structure is very similar (top vs bottom panels). This highlights the challenge of inferring source properties from structures within the solar wind as two initially different torsional waves lead to a similar ultimate switchback structure. 

\subsection{Switchback lengths scales}
In-situ studies by for example \citet{Horbury2020,Laker2021} suggest that switchbacks are usually elongated along the parallel field direction, with aspect ratios as high as $10:1$. This has previously been a challenge to reconcile with the natural tendency of Alfv\'{e}n waves to pancake as they propagate. Here we demonstrate that aspect ratios of $5$ or more appear and are maintained by the switchbacks in our simulations. To estimate parallel ($l_\parallel$) and perpendicular ($l_\perp$) lengths scales of the switchback rings as a function of radial distance we employ the following methods. For $l_\parallel$, since the switchback rings are broadly aligned to the radial direction once they form, each grid cell where $\theta_B > 90^\circ$ was identified at each time, and $l_\parallel$ is defined to be the difference between the maximum ($R_{max}$) and minimum ($R_{min}$) radial values of these points. An example is shown for a slice through the switchback ring in Case 7 in Fig. \ref{fig:len_dig}. 

For the perpendicular length scale there are a few choices. As PSP takes time series at fast tangential orbital speeds that correspond effectively to line samples through the ring, each crossing of the switchback ring would be considered as an individual switchback. We therefore choose to define $l_\perp$ as the average width of the regions where $\theta_B> 90^\circ$ in the mid-plane ($(R_{max}+R_{min})/2$). Specifically, we define a cylindrical coordinate frame centred on where the axis of the structure's rotation crosses this plane (found by an iterative method, pink dot in Fig. \ref{fig:len_dig}). We then sample the magnetic field along lines radiating from this point. $l_\parallel$ is given by the average width of the region where $\theta_B> 90^\circ$ along each sample line, with sample lines not finding anywhere above this threshold being omitted from the average. We also tested using just the maximum width sampled in this plane instead, and found similar results although in that case the aspect ratios measured were generally lower by around a factor of $2$ to $3$. 

Figure \ref{fig:multi_len} shows the results for the three groups of simulations, colour coded in the same manner as Fig. \ref{fig:theta}. {The dotted lines show the standard deviation for measurements of $\l_\perp$ (which are also applied to $l_\parallel/l_\perp$).} As the system is not perfectly cylindrically symmetric and the switchback ring is slightly tilted in the early phases, the switchback {generally} forms on one side before spreading and ultimately connecting into a full ring, e.g. Fig. \ref{fig:iso}(c). This presents as a low average $l_\perp$ at early times {and as the ring fully forms leading to reductions in the average of $l_\perp$ and a larger spread of sampled values (increased standard deviation). The measurement in $l_\parallel$ is accurate up to the grid resolution of the sampled grid ($0.048R_\odot$), which is a small percentage of the measured values, so for clarity error bars on $l_\parallel$ are omitted.} Both $l_\parallel$ and $\l_\perp$ shows steady increases with radius, while beyond the Alfv\'{e}n radius ($R_A$) the aspect ratio ($l_\parallel/l_\perp$) is generally fairly constant, lying between $\approx 5$ and $8$. This reflects the near self-similar expansion of the switchbacks once they form. 

As in the peak deflection, we find a strong dependence of the length scales on the number of turns, with a similarity between simulations where the same number of turns are injected. In Fig. \ref{fig:comb_len}(a) we show $l_\perp$ normalized by the square root of the number of turns within the initial wave. This normalization was derived empirically and most closely collapsed the curves towards each other. Cases 1-7 ($v_{max}\approx 604$\,km s$^{-1}$) are shown in red, Cases 8-14 ($v_{max}\approx 302$\,km s$^{-1}$) blue, and Cases 15-21 ($v_{max}\approx 151$\,km s$^{-1}$) green. Within the error of the measurement, the curves collapse fairly closely and follow an overall scaling in the non-linear phase of $l_\perp \propto |B|^{-1/2}$ which beyond the Alfv\'{e}n radius where $|B|\propto R^{-2}$ is equivalent to  $l_\perp \propto R$. This scaling matches what we would expect from flux conservation arguments, i.e. that $\int \mathbf{B}\cdot d\mathbf{A} \approx 2\pi|B|a^2$ should be conserved along the flux tube the wave propagates within, where here $a$ is the radius of the flux tube and $|B|$ is the average value of the field strength within the tube. In our case $l_\perp$ is approximately the difference between inner and outer cylindrically symmetric flux tubes within which the switchback ring is located, i.e. $l_\perp \approx a_{max}-a_{min}$ where $a_{max}$ is the radius of the outer tube and $a_{min}$ the radius of the inner one. Since by the argument above $a_{max}$ and $a_{min}$ both should scale as $|B|^{-1/2}$ (assuming a uniform magnetic field distribution) then so should $l_\perp$, as observed. 

Interestingly, we find an even better match when rescaling $l_\parallel$ by the number of turns, Fig. \ref{fig:comb_len}(b). This makes more intuitive sense, as if the peak deflection saturates beyond a certain number of turns as shown in Fig. \ref{fig:theta}(d) then a linear increase in turns should roughly speaking lead to a linear increase in the length of the switchback provided there is a similar deflection spread throughout the ring, as indeed can be seen from $\theta_B$ in Fig. \ref{fig:len_dig}. For $l_\parallel$ in Fig. \ref{fig:comb_len} the normalized curves match closely and follow a similar scaling of $l_{\parallel}\propto |B|^{-1/2}$ or equivalently $l_\parallel \propto R$ beyond $R_A$. Finally, Fig. \ref{fig:comb_len}(c) shows that combining the normalizations, the aspect ratio scales approximately as $l_\parallel/l_\perp \propto turns^{1/2}$.

\section{Discussion}
\label{sec:discussion}

\subsection{Theoretical considerations}
\label{sec:dis1}
In this work we have shown how torsional Alfv\'{e}n waves can evolve into switchbacks in a super-radially expanding solar wind. The basic picture is that of a steepening Alfv\'{e}n wave that induces parallel flows in its early evolution that propagate with and shear the wave as it later evolves, generating switchback field lines in-situ. The switchback itself is a mostly spherically polarised Alfv\'{e}n wave, and is supported by magnetic tension accelerating solar wind plasma upwards. 

An intuitive way to think about this wave is as a ``chain whip", with fluid elements along field lines threading the switchback acting like links in a chain dragged forward by tension. This picture is shown in Fig. \ref{fig:schem}(a). Fluid elements along a single switchback field line are shown in blue, with a particular fluid element highlighted in red. To generate the wave, fluid elements on the right section of field lines must be displaced upwards with respect to those on the left section. The wave propagates as each fluid element on the left is accelerated forward and to the right in the solar wind reference frame (white and red circles, see also the animation). In a frame of reference moving with the wave (traveling upwards at $v_a$) the fluid elements appear to flow down the field line at the Alfv\'{e}n speed (Eqn. \ref{eqn:vb}). {Since the flux tube the fluid elements reside in has a constant cross sectional width (as $|\mathbf{B}|=const.$) and as the wave is incompressible (so $\rho\approx const.$), the fluid elements must maintain a constant spacing and so will also move upwards at $v_a$ along the field line within the switchback, (cf. Fig. \ref{fig:schem}(b)). Thus} in the solar wind frame the upward motion of the plasma appears super-Alfv\'{e}nic, with the flow speed bounded by $v_a < v \leq 2v_a$ within the switchback {depending on the deflection angle}, \citep[e.g.][]{Matteini2014,Matteini2015,Longcope2025}. This is true regardless of the sign of the magnetic field, and is the reason switchbacks are always associated with an increase in the solar wind velocity ($\delta v_R>0$). This is indeed what we see once the switchback is formed (see Fig. \ref{fig:frame}). 

The role of the shear flow then also becomes clear. By displacing the fluid elements (and the magnetic field threading it) upwards through the centre of the wave as it propagates the required difference in height of the field line before and after the wave is setup. A super-radially expanding flux tube is key to this displacement forming, as shown in Fig. \ref{fig:schem}(b). In such a flux tube the flow jet is deflected by the expanding field lines, naturally generating a component of the flow across the flux surface which helps in folding the flux surface upwards. By contrast, the opposite extreme is shown in Fig. \ref{fig:schem}(c). In this case the axi-symmetry of the system means that without a change in magnetic topology \citep[e.g.][]{Shi2024} deflections of greater than 90$^\circ$ are inhibited under the action of a purely radial, axi-symmetric shear flow so as to avoid field lines crossing. 

This may in part explain why previous simulations of jet-generated torsional Alfv\'{e}n waves have failed to generate switchbacks with deflections of greater than $90^\circ$ \citep[e.g.][]{Roberts2018,Touresse2024}. In those simulations the background field is radial and expands uniformly, and so any induced flow is also radial. There is, therefore, no component of the shear flow that becomes perpendicular to the flux surface the perturbed field lines of the axi-symmetric wave reside upon. Also, in the simulations of \citet{Touresse2024} $v_{sw}+v_a$ increases before plateauing at a near constant value \citep{Touresse2026} while in these simulations $v_{sw}+v_a$ reduces with distance, Fig. \ref{fig:setup}(c). For those simulations  this also delays the steepening of the wave and the onset of non-linear effects.

Given the presence of shear flows, do we also see evidence of the Kelvin-Helmholtz Instability (KHI) in these simulations? The KHI should in theory occur in an incompressible plasma when a shear layer with $\Delta v > v_a$ is present \citep{Chandrasekhar1961}, with the condition generalising to $\Delta v > c_f$ in compressible plasmas \citep{Miura1982}, where $c_f$ is the fast mode speed. The presence of the Alfv\'{e}n wave, particularly once the deflection passes $90^\circ$ guarantees such a jump, but as discussed above the in-phase magnetic perturbation exactly balances this to propagate the wave. Fig. \ref{fig:formation} shows that the residual flow does have a strong initial shear of $\Delta v \approx 300$\,km\,s$^{-1}$ around the time the switchback forms which decays with distance. However, out to $\approx 10 R_\odot$ $\Delta v$ is less than $v_a$ (which closely matches $c_f$ in magnitude), Fig. \ref{fig:setup}(a). So we rule out the effect of the KHI in the formation of the switchbacks in these simulations.  

{Finally, for simplicity an isothermal plasma was assumed in these simulations. With a more realistic treatment of the plasma, we would still expect to obtain the same qualitative findings, that e.g. the torsional waves evolve into ring-shaped switchbacks to remain robust. There would however be heating expected from the switch-on shock \citep[e.g.][]{Orta2003}, that is assumed to be immediately lost to the system within the present isothermal assumption. This would be expected to increase the plasma temperature and in turn increase the pressure behind the shock while reducing the jump in $\mathbf{|\mathbf{B}|}$. However, we note that in our simulations the shock is weak with a Mach number only just above $1$ so this would likely be a small effect. A further altering factor is that the solar wind solution itself would also differ, requiring some form of heating to support the wind acceleration making further extrapolations of our results challenging. Clearly, further testing of our results with a full thermodynamic treatment of the plasma should be conducted in future. From the observations, whether switchbacks have an associated increase in temperature is challenging to address due to instrumentation limitations and remains an open question, \citep[see][for a discussion]{Badman2025}.}

\subsection{Relevance to observations}
Our results show that torsional Alfv\'{e}n wave pulses can evolve into switchbacks that could be measured by PSP. {This work thus joins a growing number of observational works that suggest that many switchbacks are consistent with the picture of smaller amplitude seed Alfv\'{e}n wave pulses that are amplified in-situ as they propagate and which are measured as switchbacks at PSP \citep[e.g.][]{Sioulas2026,Dai2026}.}

To further highlight this, Figs \ref{fig:probes} and \ref{fig:probes2} show examples of field lines and line samples through the switchbacks in Case 10 and 7 which were formed from wave pulses with initial turns of $\approx 1.2$ and $\approx 2.4$, respectively. The line samples assume PSP is moving rapidly through the structure so that time dependence can be ignored. The two samples firstly highlight that spherical polarisation ($|\mathbf{B}|\approx$ const.) is well maintained across both switchbacks. $|\mathbf{B}|$ is slightly elevated with respect to the background in the vicinity of the switchback, increasing and then decreasing again as the path crosses and then re-crosses the shock. For these samples $\langle|\mathbf{B}|\rangle\approx 1.3B_0$, where the average is between the two shock crossings and $B_0=|\mathbf{B}_0|$ is the value before crossing the first shock. However, increases in $|\mathbf{B}|$ can be higher or lower depending on the simulation and the radial distance of the sample (cf. Fig. \ref{fig:ht}(c)). Such compressive signatures associated with switchbacks are less common, but have been observed for some switchbacks \citep[e.g.][]{Liu2022,Larosa2021}. The samples also show that depending upon where the switchback ring is sampled single, or double switchbacks can be measured in-situ. Furthermore, when appearing as a single switchback (path A), the switchback has a characteristic asymmetry, similar to that shown in \citet{Touresse2024} but with a larger amplitude. {We note however, that these are two idealised samples taken in order to give qualitative comparisons with PSP data. For quantitative comparisons, real PSP trajectories such as those recently used in \citet{Touresse2026} taking account of the relative motion of the spacecraft to the wave should be used. Multiple passes through the different simulations would then build up a statistical picture of the in-situ signatures of the switchbacks in these simulations which could be directly compared to observations. A more detailed investigation of this aspect is beyond the scope of the present study and will be considered in future work.}

Although our simulations are focused on the behaviour of single wave pulses, and therefore can not explain (nor do they aim to) the patchy nature of switchbacks \citep[e.g.][]{Bale2021}, in \citet{Wyper2022} we showed that curtains containing multiple torsional Alfv\'{e}n waves are associated with bursty 3D interchange reconnection. {The curtain produced a patch of deflections when sampled at $4R_\odot$ which was similar to switchback patches but at a lower typical amplitude. The results presented herein suggest that were these propagated into a super-radially expanding wind that the patch would grow in amplitude, potentially producing a realistic switchback patch at PSP, but this remains to be fully tested.} In theory, swarms of torsional Alfv\'{e}n waves may also be produced by multiple jet-lets occurring for example at the base of plumes \citep[e.g.][]{Kumar2023}. A combination of multiple switchback rings originating from such clustered origins might be compatible with these observations. However, future work is needed to test how multiple switchback rings generated from torsional Alfv\'{e}n waves interact as they propagate.

We also found that the peak deflection plateaus at $\approx 130^\circ$ in our simulations, still well beyond $90^\circ$ but lower than some observed switchbacks. Beyond a threshold of $\approx 1$ turn, cases where more twist and energy was injected, resulted in longer switchbacks (with greater $l_\parallel$) rather than switchbacks with higher overall deflections. {For a given simulation, a plateau in the deflection angle is to be expected due to the reduction in $l_\parallel$ and increase in $l_\perp$ as the wave ``pancakes" \citep[e.g.][]{Volk1973}, which naturally brings the deflection back towards $90^\circ$. This is offset however by the induced shear flows, and a balance between the two effects appears to lead to the near self-similar expansion phase that we find. For a given simulation, this sets the peak deflection achieved. When more twist is injected, we find a lengthening of the switchback volume but also generally a faster shear flow (not shown). The plateau in the peak deflection beyond $\approx 1$ turn marks a shift from the effects of increasing shear flow to the elongation of the structure as further turns are injected.} {Exploring the force balance, and whether the deflection plateau in the non-linear regime can be explained using conservation of wave action or bounds based on quasi-1D waves \citep[e.g.][]{Mallet2021,Squire2022} combined with the effects of flow shear is left to future work. Future work should also} include effects such as field line curvature and pre-existing flow shear and Alfv\'{e}n speed gradients which may enable deflections as high as $180^\circ$.

A strong observational constraint is that switchbacks with deflections greater that $90^\circ$ are relatively rare below the Alfv\'{e}n surface. Some authors have actually claimed there are none at all \citep[e.g.][]{Akhavan-Tafti2024,Bandyopadhyay2025}, however care must be taken to accurately identify when the background flow is truly sub-Alfv\'{e}nic as flow speeds within switchbacks are at or above the local Alfv\'{e}n speed, Fig. \ref{fig:schem}; see e.g. \citet{Badman2025,Sioulas2026} for a discussion. Nevertheless, taking this into account switchbacks with field reversals below the Alfv\'{e}n surface are still reduced in number. This work shows a plausible mechanism for their formation and presence below the Alfv\'{e}n surface, in addition to providing in-situ signatures consistent with PSP observations beyond the Alfv\'{e}n surface also. 

{Furthermore, the switchback volume in these simulations takes the form of a high aspect ratio ring embedded within within a larger Alfv\'{e}nic perturbation. A similar conclusion was drawn recently by \citet{Li2025} from their study of two decades of WIND data.} {Another key result was the finding that} the switchbacks formed in our simulations have a close to self-similar expansion once they form. This gave switchback parallel and perpendicular length scales that increased linearly with $R$ beyond the Alfv\'{e}n radius. Such scalings are broadly consistent with switchback observations, with e.g. \citet{Mozer2021} finding the same scaling of the perpendicular length scale of $l_{\perp}\propto R$. However, their scaling of $l_\parallel \propto R^2$ is stronger than in our simulations, but ours still represent an improvement on previous Alfv\'{e}n wave models. Furthermore, we find aspect ratios in the range of $4$ to $8$ (taking account of the overestimate of the aspect ratio in the early stages of the switchback evolution), which are well within observed ranges \citep{Horbury2020}.

Finally, we also found that switchbacks developed from waves with the same overall twist profile, quantified by the peak number of turns, develop into similar switchback structures once the wave reaches the non-linear regime. This presents a challenge in diagnosing the origins of the wave as waves with the same twist profile but lower initial peak deflection can ultimately lead to effectively the same switchback observed in-situ. However, a key constraint is that the non-linear regime must be entered while the flux tube is super-radially expanding. Beyond this, as discussed in Sec. \ref{sec:dis1} the axi-symmetry of the waves likely blocks deflections of greater than $90^\circ$ from forming via the induced shear flows. Although if the axi-symmetry of the wave is broken by the coronal hole field, via e.g. background flow shear or field line curvature then this constraint is circumvented. Focusing on coronal hole fields where the axi-symmetry of the wave is maintained this means that torsional Alfv\'{e}n waves with an initially low deflection will struggle to produce switchback field reversals. In fact, Eqn. \ref{eqn:theta} can be inverted to give the linear estimate of the initial deflection angle based on the local one measured at PSP, i.e.
\begin{equation}
\theta_{B,ph}= \tan^{-1}\left( \tan(\theta_{B,PSP}) \left(\frac{B_{PSP}}{B_{ph}}\right)^{1/2}\frac{v_{PSP}+v_{a,PSP}}{(v_{a,ph}v_{a,PSP})^{1/2}}\right).
\end{equation}
providing $v_{a,ph}$ and $B_{ph}$ can be estimated via back mapping to the photosphere. This expression is valid for deflections at PSP lower than $90^\circ$, but is most accurate when $\theta_{PSP} < 60^\circ$, Fig. \ref{fig:theta}. Note that this gives an upper bound on the initial deflection angle, as non-linear effects such as flow shear and spherical polarisation can amplify smaller initial deflections to the same value measured at PSP. 

The requirement to reach non-linearity in the super-radially expanding region of the coronal hole naturally favours waves generated by more impulsive mechanisms such as those arising from interchange reconnection for producing switchback field reversals, as opposed to those generated by swirling surface motions as they are expected to have a lower initial deflection in general. Furthermore, the positive scaling of the switchback aspect ratio with number of turns initially within the wave ($l_\parallel/l_\perp \propto turns^{1/2}$) also favours impulsively generated waves such as those originating from jets and interchange reconnection. For example, torsional Afv\'{e}n waves in coronal jets have been observed to have as many as four turns \citep{Liu2019}. While torsional Alfv\'{e}n waves launched by the ejection of plasmoids during interchange reconnection can also have multiple turns and initial deflections as high as $45^\circ$  \citep{Wyper2022,Romano2025}.

\section{Summary}
\label{sec:summary}
In this work we have shown that switchbacks naturally develop from the evolution of torsional Alfv\'{e}n waves. A summary of our key results is as follows:
\begin{itemize}
    \item Switchbacks with deflections above $90^\circ$ form when the torsional Alfv\'{e}n wave reaches the non-linear regime, whereupon parallel flows are induced that feed back on the evolution of the wave, creating switchback field lines.
    \item The switchback volume takes the form of a {ring} embedded within a broader disc of inflows/outflows that support the upflow within the switchback volume. {The broader disc includes a density increase at its centre and a fast mode shock ahead of it which is associated with an increase in density/pressure and $|\mathbf{B}|$.}
    \item Switchbacks form by this mechanism in regions of super-radial flux tube expansion, i.e. below the Alfv\'{e}n point, but appear as spherically polarized Alfv\'{e}nic deflections beyond $10 R_\odot$ where PSP would encounter them.
    \item Once formed the switchbacks have an approximately self-similar expansion, with $l_\parallel \propto R$ and $l_\perp \propto R$, and aspect ratios of between $4$ and $8$ that scale with the number of turns within the wave, $l_\parallel/l_\perp \propto turns^{1/2}$.
    \item The peak deflection of the switchbacks in our diverse parametric simulations saturates at $\approx 130^\circ$, with waves initially containing more twist leading to longer switchbacks of comparable deflection. 
\end{itemize}
Collectively, our results support the idea that impulsively generated torsional Alfv\'{e}n waves associated with for example interchange reconnection, e.g. during the onset of solar jets, could evolve into switchbacks in-situ. While our results also present a challenge to less impulsively generated torsional Alfv\'{e}n waves, such as those associated with surface swirling motions. 

Due to grid limitations, our work focused on waves with spatial scales more typical of coronal jets. However, the geometric dependence upon the twist profile suggests these results can be scaled down to waves produced by smaller scale events such as those associated with e.g. jet-lets and EUV bright points. Although this should be tested in future. Furthermore, our results are also only valid when the open field contains no strong inherent asymmetries so as to maintain the axi-symmetry of the waves as they evolve. In future work we will consider the effect of field line curvature and flow and Alfv\'{e}n speed gradients, both of which are thought to enhance switchback formation \citep{Squire2022,Toth2023}. {Initial explorative tests suggest that for example when strong flow/speed gradients are present the ring-like perturbation of the field lines is maintained but becomes tilted, shifting the volume where the field lines reverse to one side of the ring and enabling deflections as high as $180^\circ$. More details will be given in a follow up publication.} {Future work should also consider the relaxation of the isothermal assumption of constant temperature, non-axisymmetric initial Alfvén wave pulses, and the interaction of multiple pulses. The latter in particular being an important step towards addressing the observed patchiness of switchbacks in the solar wind.} Finally, given the link to interchange reconnection another next step is to consider waves self-consistently generated by interchange reconnection rather than prescribed by a flow on the lower boundary. This is also the focus of ongoing work.

\section*{} 
Some of the work presented here was inspired by illuminating discussions during an International Space Science Institute (ISSI) workshop on “Magnetic Switchbacks in the Young Solar Wind” (18-22 September 2023). PW was also supported through STFC/UKRI consortium grant ST/W00108X/1, STFC small grant UKRI14137 and through a Leverhulme project grant (RPG-2023-288). The computations were sponsored by allocations on Discover at NASA's Center for Climate Simulation and on the DiRAC Data Analytic system at the University of Cambridge, operated by the University of Cambridge High Performance Computing Service on behalf of the STFC DiRAC HPC Facility (www.dirac.ac.uk) and funded by BIS National E-infrastructure capital grant (ST/K001590/1), STFC capital grants ST/H008861/1 and ST/H00887X/1, and STFC DiRAC Operations grant ST/K00333X/1. DiRAC is part of the National E-Infrastructure. Data access statement: the supporting ARMS dataset is published in Durham University’s Collection open data repository. \href{https://doi.org/10.15128/r312579s245 }{DOI: 10.15128/r312579s245}. EP and JT acknowledge financial support by the Action Thématique Soleil-Terre (ATST) of CNRS/INSU PN Astro, co-funded by CNES and CEA, and from the French national space agency (CNES) through the APR program. EP and JT also  acknowledge the Agence Nationale de la Recherche (ANR; French National Research Agency) for its financial support of the JET2SB project n°ANR-2025-CE31-7416. JT acknowledges funding from the Initiative Physique des Infinis of Sorbonne Université. 



\appendix

\section{Height-time curves}
\label{app:int}
The height-time curves in Fig. \ref{fig:ht} are created from the simulation velocities via a simple forward-Euler approach such that the radial value $\mathcal{R}(t)$ is obtained by iterating
\begin{align}
\mathcal{R}_{n+1} &=  \mathcal{R}_n + v_{model}(\mathcal{R}_n,t_n)\frac{B_R(\mathbf{x}_{fl}(\mathcal{R}_n),0)}{B(\mathbf{x}_{fl}(\mathcal{R}_n),0)}\Delta t, \\
t_{n+1} &= t_n + \Delta t,
\end{align}
where $\mathbf{x}_{fl}(R)=(R,\theta(r),\phi(R))$ is the curve defined by the unperturbed field line at $t=0$ and $\mathcal{R}_0$ and $t_0$ are specified based on what the curve is being compared to. The magnetic field factors project the path on to the radial direction, i.e. they account for when the magnetic field is not radial. Two velocity models are considered;
\begin{align}
v_{model}(\mathcal{R}_n,t_n) = v_{sw}(\mathbf{x}_{fl}(\mathcal{R}_n),0)+v_a(\mathbf{x}_{fl}(\mathcal{R}_n),0),
\end{align}
which gives the propagation speed of a linear Alfvén wave based on the unperturbed background solar wind and Alfvén speeds, and
\begin{align}
v_{model}(\mathcal{R}_n,t_n) = v_{sw}(\mathbf{x}_{fl}(\mathcal{R}_n),0)+ {v_{a}(\mathbf{x}_{fl}(\mathcal{R}_n),t_n)},
\end{align}
which uses the time-evolving {Alfv\'{e}n speed, $v_a(\mathbf{x}_{fl}(\mathcal{R}_n),t_n)$}.

\section{WKB derivation of Alfv\'enic amplitudes in an expanding solar-wind flux tube}
\label{app:theta}

Consider a nearly radial open magnetic flux tube with background radial field
\(B_0(R)\), density \(\rho(R)\), and outflow speed \(v_{sw}(R)\).  The Alfv\'en speed is
\begin{equation}
  v_a(R) \equiv \frac{B_0}{\sqrt{4\pi \rho}} ,
\end{equation}
and we define the Alfv\'en Mach number
\begin{equation}
  M_{\rm A}(R) \equiv \frac{v_{sw}}{v_a} ,
\end{equation}
so that the Alfv\'en critical point \(R = R_{\rm A}\) is the location where \(M_{\rm A}=1\).
It is also convenient to introduce
\begin{equation}
  \eta(R) \equiv \frac{\rho(R)}{\rho_{\rm A}} ,
\end{equation}
where \(\rho_{\rm A} \equiv \rho(R_{\rm A})\) is the density at the Alfv\'en point.

We assume a thin flux tube whose cross-sectional radius \(a(R)\) varies with
\(R\). Conservation of magnetic flux and mass flux implies
\begin{equation}
  B_0 a^2 = \text{const}, \qquad
  \rho v_{sw} a^2 = \text{const},
\end{equation}
so that
\begin{equation}
  B_0 \propto a^{-2}, \qquad
  \rho \propto \frac{1}{v_{sw} a^2} .
\end{equation}
Combining these with the definition of \(v_a\) gives
\begin{equation}
  v_a \propto a^{-1} v_{sw}^{1/2}, \qquad
  v_{sw} \propto a^2 v_a^2, \qquad
  M_{\rm A} = \frac{v_{sw}}{v_a} \propto a^2 v_a.
  \label{eq:UBa-relations}
\end{equation}
This also gives
\begin{equation}
  M_{\rm A} = \eta^{-1/2}, \qquad
  v_a = \eta^{1/2} v_{sw} .
\end{equation}

\paragraph{WKB framework and wave-action variables}

We consider outward-propagating Alfv\'en waves in the WKB limit, in which the
parallel wavelength is small compared to the scale height of the background:
\begin{equation}
  k_\parallel L_{\rm bg} \gg 1.
\end{equation}
In this limit the wave frequency in the plasma frame is approximately conserved
as the wave propagates outwards, and the appropriate adiabatic invariant is the
wave action (energy divided by frequency).  For Alfv\'enic fluctuations in a
radially stratified flow, the WKB evolution can be cast in terms of the
Heinemann-Olbert variables 
\citep{Heinemann1980},
\begin{equation}
  f^+ = \frac{1+\eta^{1/2}}{\eta^{1/4}}\, z^+, \qquad
  f^- = \frac{1-\eta^{1/2}}{\eta^{1/4}}\, z^- ,
  \label{eq:HO-wave-action}
\end{equation}
as extended and discussed
in \citet{Chandran2009,Chandran2019}, where \(\boldsymbol{z}^{\pm} = \bm{v} \mp \bm{b}/\sqrt{4\pi \rho}\) are the outward- and inward-propagating Elsasser fields.
The Heinemann-Olbert variables, which are the square root of the 
wave action, remain constant if 
 reflection and nonlinear
interactions are neglected, meaning that the outward-propagating Elsasser field satisfies 
\begin{equation}
  z^+ \propto  \frac{\eta^{1/4}}{1+\eta^{1/2}} = \left(M_{\rm A}^{1/2}+M_{\rm A}^{-1/2}\right)^{-1}.
  \label{eq:HO-wave-action}
\end{equation}

For an outward Alfv\'en wave with negligible inward component
(\(z^- \approx 0\)), the transverse velocity and magnetic-field perturbations
satisfy
\begin{equation}
  z^+ \simeq 2 v_\perp \simeq 2 \frac{\delta B_\perp}{\sqrt{4\pi \rho}} ,
  \label{eq:chi-plus-MA-again}
\end{equation}
so up to a factor of order unity the WKB scaling of \(v_\perp\) is the same as
that of \(z^+\).  In particular,
\begin{equation}
  \frac{v_\perp}{v_a} \;\propto\; \frac{z^+}{v_a} .
\end{equation}
Below we will therefore quote results in terms of \(v_\perp/v_a\), with
the understanding that factors of two are unimportant for the scaling.

\paragraph{General expression for $v_\perp/v_a$.}

Let $R_{\rm ph}$ denote a reference radius (e.g.\ the photospheric or
lower-coronal boundary used in the simulations), and denote
\begin{equation}
  M_{{\rm A},\rm ph} \equiv M_{\rm A}(R_{\rm ph}), \qquad
  z^+_{\rm ph} \equiv z^+(R_{\rm ph}), \qquad
  v_{a,{\rm ph}} \equiv v_a(R_{\rm ph}).
\end{equation}
Evaluating \eqref{eq:chi-plus-MA-again} at $R$ and at $R_{\rm ph}$ and equating
the two expressions,
\begin{equation}
  \bigl(M_{\rm A}^{1/2}+M_{\rm A}^{-1/2}\bigr) z^+(R)
  = \bigl(M_{{\rm A},\rm ph}^{1/2}+M_{{\rm A},\rm ph}^{-1/2}\bigr) z^+_{\rm ph},
\end{equation}
we obtain the general WKB scaling of the Elsasser amplitude:
\begin{equation}
  z^+(R) =
  z^+_{\rm ph}\,
  \frac{M_{{\rm A},\rm ph}^{1/2}+M_{{\rm A},\rm ph}^{-1/2}}%
       {M_{\rm A}^{1/2}+M_{\rm A}^{-1/2}}.
  \label{eq:zplus-of-R-ph}
\end{equation}
This  yields the amplitude 
expression
\begin{equation}
  \frac{v_\perp}{v_a}(R)
  \;\simeq\; \frac{v_{\perp,{\rm ph}}}{v_{a,{\rm ph}}}\,
  \frac{v_{a,{\rm ph}}}{v_a(R)}\,
  \frac{M_{{\rm A},\rm ph}^{1/2}+M_{{\rm A},\rm ph}^{-1/2}}%
       {M_{\rm A}(R)^{1/2}+M_{\rm A}(R)^{-1/2}}.
  \label{eq:vperp-over-vA-general-ph}
\end{equation}
This is the WKB prediction for how the dimensionless amplitude $v_\perp/v_a$
should vary with radius, given the background profiles $v_{sw}(R)$,
$v_a(R)$, and hence $M_{\rm A}(R)=v_{sw}/v_a$ used in the simulation.

\paragraph{Linear field-line deflection angle.}
The local field-line deflection angle is
\begin{equation}
  \tan\theta = \frac{B_\perp}{B_x},
\end{equation}
where $B_x$ is the component of the field along the background direction. If we assume that $B_\perp$ grows without inducing a parallel perturbation then $B_x\simeq B_0$ and
\begin{equation}
  \tan\theta_{\rm lin} \simeq \frac{B_\perp}{B_0}.
\end{equation}
In practice, this is an underestimate as any induced parallel component will generate a larger deflection, however, it serves as a useful lower bound. For Alfv\'enic fluctuations,
\begin{equation}
  \frac{B_\perp}{\sqrt{4\pi\rho}} \simeq v_\perp
  \quad\Rightarrow\quad
  \frac{B_\perp}{B_0} \simeq \frac{v_\perp}{v_a},
\end{equation}
So under these assumptions a lower bound for the deflection is given by
\begin{equation}
  \theta_{lin}
  \;\simeq\; \tan^{-1}\left(\tan(\theta_{ph})\,
  \frac{v_{a,{\rm ph}}}{v_a(R)}\,
  \frac{M_{{\rm A},\rm ph}^{1/2}+M_{{\rm A},\rm ph}^{-1/2}}{M_{\rm A}(R)^{1/2}+M_{\rm A}(R)^{-1/2}} \right).
  \label{eqn:app_th}
\end{equation}
This can be re-expressed in terms of the primary variables of the solar wind flow by noting that
\begin{align}
    M_{{\rm A}}^{1/2}+M_{{\rm A}}^{-1/2}=M_{{\rm A}}^{-1/2}v_a^{-1}(v_{sw}+v_a)
\end{align}
and that
\begin{align}    
    M_A B/v_a &= M_{A,{\rm ph}}B_{\rm ph}/v_{a,{\rm ph}}\\
    \implies M_A/M_{A,{\rm ph}} &= (B_{\rm ph}/B)(v_a/v_{a,{\rm ph}})
\end{align}
which when substituted into Eqn. \ref{eqn:app_th} gives
\begin{equation}
  \theta_{lin}
  \;\simeq\; \tan^{-1}\left(\tan(\theta_{ph})\,
  \left(\frac{v_a(R)}{v_{a,{\rm ph}}}\right)^{1/2}\,
  \left(\frac{B_{\rm ph}}{B(R)}\right)^{1/2}\,
  \frac{v_{{sw, \rm ph}}+v_{a,{\rm ph}}}{v_{sw}(R)+v_a(R)} \right).
\end{equation}
This expression gives a lower estimate of the maximum field line deflection within the Alfvén wave, neglecting non-linear effects. Since this expression applies locally to the most maximally deflected field line within the wave it also is true for any Alfv\'{e}n wave with the same initial peak magnetic deflection, regardless of e.g. wave energy, number of turns or wave pulse length scales.


\bibliography{biblionew}
\bibliographystyle{aasjournal}



\end{document}